\documentclass[aps,article,twocolumn,superscriptaddress,floatfix]{revtex4-1}

\usepackage{mathtools}
\usepackage[usenames,dvipsnames]{xcolor}
\usepackage{float}
\usepackage{amssymb}
\usepackage{bm}
\usepackage{tabularx, booktabs}
\newcolumntype{Y}{>{\centering\arraybackslash}X}

\definecolor{dgreen}{rgb}{0,.5,0}
\definecolor{dblue}{rgb}{0,0,.5}
\definecolor{dred}{rgb}{0.5,0,.5}

\begin{document}

\title{Site-Occupation Embedding Theory using Bethe Ansatz Local Density Approximations}

\author{Bruno Senjean}
\thanks{Corresponding author}
\email{senjean@unistra.fr}
\affiliation{Laboratoire de Chimie Quantique,
Institut de Chimie, CNRS/Universit\'{e} de Strasbourg,
4 rue Blaise Pascal, 67000 Strasbourg, France}

\author{Naoki Nakatani}
\affiliation{Department of Chemistry,
Graduate School of Science and Engineering,
Tokyo Metropolitan University,
1-1 Minami-Osawa, Hachioji,
Tokyo 192-0397, Japan}
\affiliation{Institute for Catalysis,
Hokkaido University,
N21W10 Kita-ku, Sapporo,
Hokkaido 001-0021, Japan}

\author{Masahisa Tsuchiizu}
\affiliation{Department of Physics,
Nara Women's University,
Nara 630-8506, Japan}

\author{Emmanuel Fromager}
\affiliation{Laboratoire de Chimie Quantique,
Institut de Chimie, CNRS/Universit\'{e} de Strasbourg,
4 rue Blaise Pascal, 67000 Strasbourg, France}

\date{\today}

\begin{abstract}

Site-occupation embedding theory (SOET) is an alternative 
formulation of density-functional theory (DFT) for model
Hamiltonians where
the fully-interacting Hubbard problem is mapped, in principle exactly, 
onto an
impurity-interacting (rather than a non-interacting) one. It provides a 
rigorous framework for combining
wavefunction (or Green function) based methods with DFT.
In this work, exact expressions for the per-site energy and double
occupation of the uniform Hubbard model are derived in the context of
SOET. As readily seen from these
derivations, the so-called bath contribution to the per-site correlation energy
is, in addition to the latter, the key density-functional quantity to
model in SOET. Various approximations based on Bethe ansatz and
perturbative solutions to the Hubbard and single impurity Anderson
models are constructed and tested on a 
one-dimensional
ring. The self-consistent
calculation of the embedded impurity wavefunction has been performed
with the density matrix renormalization group
method. 
It has been shown that
promising results are obtained in specific regimes of correlation
and density.
Possible further developments have been proposed  in order
to provide reliable embedding functionals and potentials. 

\end{abstract}

\maketitle

\section{Introduction}

Although the independent-particle picture is applicable to 
many electronic systems, such as conventional metals and band 
insulators, it drastically fails when electron correlation 
becomes strong, like in transition metal oxides 
where metal-insulator transitions occur.
Describing such a transition accurately at the
computational cost of an independent-particle theory is
still a challenge. In the context of density-functional theory (DFT), a correction based on the on-site two-electron repulsion
parameter $U$ can be explicitly added
to the exchange-correlation functional, in the spirit of hybrid functionals, 
thus leading to the so-called DFT+$U$ method~
\cite{anisimov1991band,liechtenstein1995density}.
But still, some crucial aspects are missing in the DFT+$U$, for instance strongly-correlated phenomena such as the Kondo effect, which
cannot be treated within a static mean-field approximation.
Therefore, it is necessary to consider a many-body picture
of the problem.

Because it often appears that the region of interest is only one part of a much larger system
and considering strong electron correlation as essentially 
local~\cite{pulay1983localizability, saebo1993local,hampel1996local},
embedding approaches 
are mainly used in practice~\cite{sun2016quantum}.
In these approaches, the whole system is usually mapped onto an embedded quantum problem, e.g. a small system called impurity and the rest of the system called the bath~\cite{ayral2017dynamical}.
The {\it dynamical mean-field theory}
(DMFT)~\cite{georges1992hubbard,georges1996limitdimension,
kotliar2004strongly,held2007electronic,zgid2011DMFTquantum} 
has been proved to treat efficiently systems with 
$d$ or $f$ localized shells,
%
however, there are still some cases for which DMFT is not sufficiently
accurate, 
especially in the case 
when non local electron correlation becomes important.
In order to further improve on its
performance, combined
DMFT+DFT~\cite{kotliar2006reviewDMFT} or
DMFT+GW~\cite{sun2002extended,biermann2003first,karlsson2005self,
boehnke2016strong,werner2016dynamical,nilsson2017multitier} 
schemes have been proposed to 
recover such effects.
In another promising approach, 
the so-called {\it self-energy embedding theory}
\cite{kananenka2015systematically,lan2015communication,
lan2017testing}, strong 
correlation is not considered as strictly local, which can be
appreciable for real compounds.
Its applicability to both model and {\it ab-initio} Hamiltonians is also
appealing.
All these embedding techniques 
are formulated in terms of the (frequency-dependent) one-particle 
Green function. On the other hand, in the {\it density-matrix embedding theory
} (DMET)~\cite{knizia2012density,knizia2013density,bulik2014density,zheng2016ground,wouters2016practical,
wouters2016five,rubin2016hybrid}, the 
embedded fragment (impurity) is described with a high-level wavefunction-based method 
while the rest of the system is usually treated at the mean-field level.
Extensions have been 
proposed in order to include correlation in the 
bath~\cite{tsuchimochi2015density} or 
for improving the description of the boundary between the fragment and 
the bath~\cite{welborn2016bootstrap}.

Turning to DFT, its extension to model Hamiltonians is usually 
referred to as {\it site-occupation functional theory}
(SOFT)~\cite{chayes1985density,gunnarsson1986density,DFT_lattice,DFT_ModelHamiltonians}.
In conventional Kohn-Sham (KS) SOFT, the physical fully-interacting many-body 
problem is mapped onto a noninteracting one by means of a
Hartree-exchange-correlation (Hxc) functional of the density 
(
i.e.,
 the sites occupation in this context).
The SOFT has been shown to give very accurate density and energy profiles with the 
{\it Bethe ansatz local density approximation} (BALDA)~
\cite{lima2002density,lima2003density,capelle2003density}, the spin-dependent BALDA~
\cite{franca2012simple}, and its fully numerical formulation~
\cite{xianlong2006bethe,xianlong2007luther}.
The methods have been applied to both repulsive 
~\cite{silva2005effects,akande2010electric}
and attractive
~\cite{campo2005phase,xianlong2008effects} Hubbard models.
The electronic transport has also been studied by applying SOFT to the
Anderson junction model~
\cite{bergfield2012bethe,liu2012accuracy,liu2015coulomb}.
Time-dependent
\cite{verdozzi2008time,kurth2010dynamical,karlsson2011time} and
temperature-dependent
extensions~\cite{stefanucci2011towards,xianlong2012lattice} have been
investigated over the years. Other reduced quantities can also be used, as first discussed by 
Sch\"onhammer {\it et al.}~\cite{DFT_lattice}, such as the 1-body 
reduced density matrix~
\cite{schindlmayr1995density,lopez2002density,lopez2004interaction,
saubanere2009scaling,
saubanere2011density,tows2011lattice,
tows2014density,
saubanere2016interaction}, or the steady current in connection with
steady-state transport~\cite{stefanucci2015steady,kurth2016nonequilibrium}.

For the purpose of modelling strongly correlated regimes, some of the authors
have recently proposed an alternative formulation of 
SOFT where, in contrast to standard KS SOFT, the interacting Hubbard
problem is mapped onto an impurity-interacting one, 
thus leading to an in-principle-exact {\it site-occupation embedding 
theory} (SOET)~\cite{fromager2015exact,senjean2017local}. 
In order to turn SOET into a practical computational method, embedding
density-functional approximations must be developed. This has been done
so far only for the asymmetric Hubbard dimer~\cite{senjean2017local}. In this work we show how
Bethe ansatz and perturbative solutions to both Hubbard and Anderson models can
be used for designing local density approximations in the context of
SOET. In Ref.~\cite{senjean2017local}, the self-consistent impurity
problem SOET relies on
was solved for a 8-site model by exact diagonalization, as a proof of
concept. In this work, we also present an implementation of SOET where
the embedded impurity system is treated with the density matrix renormalization group (DMRG) method~
\cite{white1992density,white1993density,verstraete2008matrix,
schollwock2011density,naokicode}, thus allowing for calculations on
larger rings.

The paper is organized as follows. After a brief review of SOET
(Sec.~\ref{subsec:SOET}), exact SOET-based expressions for the per-site energy and
double occupation are derived in Sec.~\ref{subsec:exact_expressions}, in the particular case of the uniform one-dimensional Hubbard
problem. Exact properties of the embedding functionals are also
presented. The construction of local density-functional approximations
is then discussed in Sec.~\ref{subsec:DFA}. A
connection between SOET and the single impurity Anderson model
is investigated in Sec.~\ref{subsec:Hubbard_SIAM}. A summary of the
various approximations tested in this work 
is given in Sec.~\ref{sec:Computational details}
with the computational
details. The
results are discussed in Sec.~\ref{sec:results}. Conclusions and
perspectives are finally given in Sec.~\ref{sec:conclusion}.

\section{Theory}\label{sec:theory}

\subsection{Site occupation embedding theory}\label{subsec:SOET}

We focus on the one-dimensional Hubbard model in an external potential
$\mathbf{v} \equiv \lbrace v_i \rbrace_i$,
\begin{eqnarray}\label{eq:Hubb_hamiltonian}
\hat{H}(\mathbf{v}) = \hat{T} + \hat{U} + \hat{V},
\end{eqnarray}
where the hopping operator,
\begin{eqnarray}
\hat{T} = -t 
\sum_{i=0}^{L-1} 
\sum_{\sigma=\uparrow,\downarrow} 
\left(
\hat{c}^\dagger_{i\sigma}\hat{c}_{i+1\sigma}
+\mathrm{H.c.}
\right)
,
\end{eqnarray}
is the analog of the kinetic 
energy operator in {\it ab-initio} Hamiltonians.
Here 
$t$ is the hopping integral and 
$\hat c_{i\sigma}^\dagger$ is the creation operator of an electron
 at the $i$th site with spin $\sigma=\uparrow,\downarrow$.
The site index $i$ runs from $0$ to $L-1$ where 
 $L$ is the number of sites, and we impose the 
 periodic boundary condition
$\hat c_{L\sigma}=\hat c_{0\sigma}$.
The on-site two-electron repulsion operator is given by
\begin{eqnarray}
\hat{U}  =   U 
\sum_{i=0}^{L-1} \hat{n}_{i\uparrow} \, \hat{n}_{i\downarrow}
,
\end{eqnarray}
where 
$\hat{n}_{i\sigma} = 
\hat{c}_{i\sigma}^\dagger \hat{c}_{i\sigma}$
and $U$ represents the Hubbard interaction between spin-up and
spin-down electrons on the same
site.
In order to formulate the SOFT and SOET, we further introduce
the on-site potential operator by 
\begin{eqnarray}
\hat{V}  =  \sum_{i=0}^{L-1} v_i \, \hat{n}_i
,
\end{eqnarray}
where
$\hat{n}_i = \hat{n}_{i\uparrow} + \hat{n}_{i\downarrow}$.
In order to have a self-contained paper, this subsection summarizes the
main equations of Ref.~\cite{senjean2017local}.

In SOFT, the exact ground-state energy is obtained variationally 
as
\begin{eqnarray}
E(\mathbf{v})= \underset{\mathbf{n}}{\rm min} \left \lbrace F(\mathbf{n}) + 
(\mathbf{v}| \mathbf{n}) \right \rbrace, 
\label{eq:E=F+v}
\end{eqnarray}
where $\mathbf{n} \equiv \lbrace n_i \rbrace_i$ is the site-occupation
(simply called 
density in the following)
vector
and
$(\mathbf{v}|
\mathbf{n}) = \sum_i v_in_i$.
The Levy--Lieb (LL) functional reads
\begin{eqnarray}
F(\mathbf{n}) = \underset{\Psi \rightarrow \mathbf{n}}{\rm min} \left 
\lbrace \langle \Psi \vert \hat{T} + \hat{U} \vert \Psi \rangle \right 
\rbrace
,
 \label{eq:HK}
\end{eqnarray}
where the minimization is restricted to wavefunctions $\Psi$ with density
$\mathbf{n}$.

In the conventional KS formalism,
the LL functional is decomposed as
\begin{eqnarray}
F(\mathbf{n}) = T_{\rm s}(\mathbf{n}) + E_{\rm Hxc}
(\mathbf{n})
,
\end{eqnarray}
where
$T_{\rm s}(\mathbf{n}) = \underset{\Psi \rightarrow 
\mathbf{n}}{\rm min}  \{ \langle \Psi \vert \hat{T}\vert \Psi 
\rangle \}$
is the $t$-dependent analog of the noninteracting kinetic 
energy functional and 
\begin{eqnarray}
E_{\rm Hxc}(\mathbf{n})=\dfrac{U}{4}\sum_in_i^2+E_{\rm c}(\mathbf{n})
\end{eqnarray}
is the 
$t$- and $U$-dependent Hxc 
functional. The latter is ``universal'' in a sense that it 
does not depend on the external potential $\mathbf{v}$.

Turning
to the SOET~\cite{fromager2015exact,senjean2017local}, 
we label, for convenience,  the location of the ``to be embedded'' impurity site in 
real (discretized) space as $i=0$. 
The LL functional 
is then decomposed into impurity
and bath contributions as
\begin{eqnarray}\label{eq:HK_decomposition}
F(\mathbf{n}) = F^{\rm imp}(\mathbf{n}) + \overline{E}_{\rm Hxc}^{\rm 
bath}(\mathbf{n}),
\end{eqnarray}
where the impurity-interacting LL functional reads
\begin{eqnarray}
F^{\rm imp}(\mathbf{n}) &=&  \underset{\Psi \rightarrow \mathbf{n}}
{\rm min} \left \lbrace \langle \Psi \vert \hat{T} + \hat{U}_0 \vert \Psi 
\rangle \right \rbrace
,
 \label{eq:HK_imp}
\end{eqnarray}
with $\hat{U}_0=U \hat{n}_{0\uparrow}\hat{n}_{0\downarrow}$. 
By using Eqs.~(\ref{eq:E=F+v}), (\ref{eq:HK_decomposition}), and 
(\ref{eq:HK_imp}),
we obtain 
the exact 
SOET energy expression~\cite{fromager2015exact,senjean2017local},
\begin{equation}
\label{eq:variational_energy}
E(\mathbf{v}) = \underset{\Psi}{\rm
min}\left\lbrace 
\langle\Psi\vert\hat{T}+ \hat{U}_0 \vert\Psi\rangle +  \overline{E}^{\rm 
bath}_{\rm Hxc}(\mathbf{n}^{\Psi}) + (\mathbf{v}| \mathbf{n}^{\rm \Psi})
 \right\rbrace
,
 \end{equation}
where $\mathbf{n}^{\rm
\Psi}\equiv\lbrace\langle\Psi\vert\hat{n}_i\vert\Psi\rangle\rbrace_i$ 
is the density of the trial many-body wavefunction $\Psi$. 
The 
optimized
impurity-interacting wavefunction
$\Psi^{\rm imp}$ in Eq.~(\ref{eq:variational_energy})
fulfills the following self-consistent equation,
\begingroup
\begin{eqnarray}\label{eq:self-consistent-SOET}
&&\displaystyle \left( \hat{T} + \hat{U}_0 + \displaystyle \sum_{i}  \left[ v_i 
+
 \dfrac{\partial \overline{E}^{\rm bath}_{\rm Hxc}(\mathbf{n}^{\Psi^{\rm 
 imp}})}{\partial n_i} 
\right]
 \hat{n}_i \right) \vert \Psi^{\rm imp} \rangle 
\nonumber \\
&& = \mathcal{E}^{\rm imp} \vert \Psi^{\rm imp} \rangle , 
\end{eqnarray}
\endgroup
where $\lbrace{v_i+\partial  \overline{E}^{\rm bath}_{\rm
Hxc}(\mathbf{n}^{\Psi^{\rm imp}})/\partial n_i\rbrace}_i$ plays 
the role of an embedding
potential for the impurity. This potential is unique (up to a constant)
and ensures, like the KS potential in
conventional DFT, that $\Psi^{\rm
imp}$ reproduces the exact ground-state density of the true
(fully-interacting) Hubbard
Hamiltonian.
Any correlated method based on the explicit calculation
of many-body wavefunctions or Green functions could in principle be
applied 
for solving Eq.~(\ref{eq:self-consistent-SOET}).
Obviously, in order to perform
practical SOET calculations, it is necessary to develop 
approximations to the complementary bath functional introduced in
Eq.~(\ref{eq:HK_decomposition}). This is the main focus of this paper.
Let us first consider the following KS decomposition of Eq.~(\ref{eq:HK_imp}),
\begin{eqnarray}
F^{\rm imp}(\mathbf{n}) = T_{\rm s}(\mathbf{n}) + E^{\rm imp}_{\rm Hxc}
(\mathbf{n}), \label{eq:HK_KS_imp}
\end{eqnarray}
where 
\begin{eqnarray}
E^{\rm imp}_{\rm Hxc}(\mathbf{n})= \dfrac{U}{4} n_0^2+E^{\rm imp}_{\rm
c}(\mathbf{n})
\label{eq:Hx_plus_c_for_imp}
\end{eqnarray}
is the analog of the Hxc functional for the
impurity-interacting system.
By combining Eqs.~(\ref{eq:HK_decomposition}) and 
(\ref{eq:HK_KS_imp}), we have
\begin{eqnarray}\label{eq:decomp_Hx_c_bath}
\overline{E}_{\rm Hxc}^{\rm bath}(\mathbf{n}) = \dfrac{U}{4}\sum_{i\neq 0} n_i^2 +  \overline{E}_{\rm c}^{\rm bath}(\mathbf{n}), 
\end{eqnarray}
where the exact correlation functional for the bath,
\begin{eqnarray}
\label{eq:Ecbath_expression}
\overline{E}_{\rm c}^{\rm bath}(\mathbf{n})=E_{\rm c}(\mathbf{n}) - E_{\rm 
c}^{\rm imp} (\mathbf{n}),
\end{eqnarray}
is simply the
difference in correlation energy between the fully-interacting system
(i.e.,
an interacting impurity site surrounded by interacting bath
sites), for which local density-functional approximations have already been
developed (see, for example,
Refs.~\cite{lima2003density,xianlong2012lattice}) and the auxiliary system consisting of an interacting impurity
site surrounded by non-interacting bath sites,
{\it both systems having the same density} $\mathbf{n}$. In the rest of
this work we will discuss various strategies for developing local density
functional approximations to $\overline{E}_{\rm c}^{\rm
bath}(\mathbf{n})$ or, equivalently, $E_{\rm c}(\mathbf{n})$ {\it and} $E_{\rm 
c}^{\rm imp} (\mathbf{n})$. For that purpose, we will first derive in the next section exact properties of the latter functionals
for a uniform system.

\subsection{Exact SOET for the uniform Hubbard model}\label{subsec:exact_expressions}

Following Capelle and coworkers~\cite{lima2002density}, we will use as 
reference the
uniform Hubbard system ($\mathbf{v} =\mathbf{0}$) in the following 
in order to derive  
local density
approximations for $\overline{E}_{\rm c}^{\rm bath}(\mathbf{n})$. 
In this context, the standard density-functional correlation
energy~\cite{lima2003density},
\begin{eqnarray}
E_{\rm c}(\mathbf{n}) = \sum_i e_{\rm c}(n_i),
\end{eqnarray}
is simply expressed in terms of the
per-site correlation energy $e_{\rm c}(n)$     
for which an exact analytical expression can be obtained at half-filling from the Bethe ansatz~\cite{NoMott_Hubbardmodel}. 
For convenience, we introduce the per-site analog of
Eq.~(\ref{eq:Ecbath_expression}), 
\begin{eqnarray}\label{eq:local_ecbath_exp}
\overline{e}_{\rm c}^{\rm
bath}(\mathbf{n})=e_{\rm c}(n_0)- E_{\rm c}^{\rm imp}(\mathbf{n}),
\end{eqnarray}
which
leads
to the final expression,
\begin{eqnarray}\label{eq:final_exp_Ec_bath}
\overline{E}_{\rm c}^{\rm bath}(\mathbf{n})
& = & \sum_{i\neq 0} e_{\rm c}(n_i) + 
\overline{e}_{\rm c}^{\rm bath}(\mathbf{n}).
\end{eqnarray}
Let us stress that the deviation of the impurity correlation 
energy $E_{\rm c}^{\rm imp}(\mathbf{n})$ from the conventional (total) per-site correlation energy $e_{\rm c}(n_0)$ is the key
density-functional quantity to
model in the SOET. It becomes
even more clear when considering, for example, the exact (uniform) double site-occupation expression~\cite{DFT_ModelHamiltonians}
\begin{eqnarray}
d=\langle\hat{n}_{i\uparrow}\hat{n}_{i\downarrow}\rangle=\dfrac{1}{L}\dfrac{\partial E}{\partial U}=\dfrac{n^2}{4}+\dfrac{\partial e_{\rm
c}(n)}{\partial U}, \label{eq:dblocc_KS}
\end{eqnarray}   
where
$n$ denotes the uniform density in the reference Hubbard system
with total energy $E=E({\mathbf{v}=\mathbf{0}})$. 
As shown in Appendix~\ref{sec:dblocc}, the following equivalent expression is obtained in SOET,
\begin{eqnarray}\label{eq:doubleocc_soet}
d& = & d^{\rm imp} + \dfrac{\partial 
\overline{e}_{\rm c}^{\rm bath}(\mathbf{n}^{\Psi^{\rm imp}})}{\partial U} ,
\end{eqnarray}
where $d^{\rm imp} = \langle \Psi^{\rm imp} \vert 
\hat{n}_{0\uparrow}\hat{n}_{0\downarrow}  \vert\Psi^{\rm imp} \rangle $ 
is the double occupation of the impurity site for the
impurity-interacting wavefunction $\Psi^{\rm imp}$. Note that, in
the exact theory, 
the latter is expected to reproduce 
the uniform density 
only (i.e.,
$\underline{n}\equiv\big\{n_i=n\big\}_i\equiv\mathbf{n}^{\Psi^{\rm imp}}$) and {\it not} the double occupation, hence the second
density-functional contribution on the right-hand side of
Eq.~(\ref{eq:doubleocc_soet}). Turning to the per-site
energy~\cite{lima2003density}, 
\begin{eqnarray}\label{eq:per-site-energy_KS}
e=\frac{E}{L}=t_{\rm s}(n)+\dfrac{U}{4}n^2+e_{\rm c}(n),
\end{eqnarray}   
where $t_{\rm s}(n)=-4t\sin(\pi n/2)/\pi$ in the thermodynamic limit
($L\rightarrow+\infty$), we
equivalently obtain in   
SOET (see the proof in Appendix~\ref{sec:persite}) the following {\it
exact} expression,
\begin{eqnarray}
e &=& t_{\rm s}(n^{\Psi^{\rm imp}}_0)+U d^{\rm imp}  
+t \dfrac{\partial e_{\rm c}(n^{\Psi^{\rm imp}}_0)}{\partial t} 
+ \overline{e}_{\rm c}^{\rm bath}(\mathbf{n}^{\Psi^{\rm imp}})
\nonumber\\
&&
 - t\dfrac{\partial 
\overline{e}_{\rm c}^{\rm bath}(\mathbf{n}^{\Psi^{\rm imp}})}{\partial t}.
\label{eq:per-site-energy}
\end{eqnarray}
In contrast to the regular KS expressions [Eqs.~(\ref{eq:dblocc_KS})
and (\ref{eq:per-site-energy_KS})],
our SOET expressions [Eqs.~(\ref{eq:doubleocc_soet}) and (\ref{eq:per-site-energy})]
involve the (embedded impurity) double occupation explicitly, which can improve on the results
significantly when
approximate density functionals are used, as shown in
Sec.~\ref{sec:results}.

Returning to the exact theory, let us
now highlight
some properties of 
$\overline{e}_{\rm c}^{\rm bath}(\mathbf{n})$. Since the standard 
per-site correlation  
functional as well as the impurity one
are invariant under hole-particle
symmetry (see Ref.~\cite{DFT_ModelHamiltonians} and Appendix~\ref{sec:particle-hole}), the  
per-site bath correlation functional is also invariant, according to
Eq.~(\ref{eq:local_ecbath_exp}),
i.e. $\overline{e}_{\rm c}^{\rm bath}(\mathbf{n}) = \overline{e}_{\rm c}^{\rm
bath}
(\underline{2}-\mathbf{n})$.
Consequently, the exact embedding potential in the uniform case 
[see Eq.~(\ref{eq:self-consistent-SOET})], 
\begin{eqnarray}\label{eq:emb_pot_uniform}
\left. \dfrac{\partial \overline{E}^{\rm bath}_{\rm Hxc}(
\mathbf{n}
)}{\partial n_i}\right|_{\mathbf{n}=\underline{n}}
&=&(1-\delta_{i0})\left(\dfrac{Un}{2}+\dfrac{\partial
e_{\rm c}(n)}{\partial n}\right)
\nonumber
\\
&&{}
+\left.\dfrac{\partial \overline{e}_{\rm
c}^{\rm bath}(\mathbf{n})}{\partial
n_i}\right|_{\mathbf{n}=\underline{n}} 
,
\end{eqnarray}
will, at half-filling ($n = 1$) and for a {\it finite-size} system, be equal to 
$U/2$ everywhere in the bath and zero on the impurity site or,
equivalently, $-U/2$ on the impurity site and zero in the
bath (see Appendix~\ref{sec:particle-hole}). In this
particular case, the auxiliary impurity-interacting system is similar to the
symmetric single-impurity Anderson model 
(SIAM)~\cite{Anderson}. This feature has already been
observed numerically in the
particular case of a 8-site ring in Ref.~\cite{senjean2017local} but, at
the time, it was not rationalized in terms of
hole-particle symmetry as we just did. 
Away from half-filling, 
the embedding potential loses its uniformity in the
bath since the interaction on the impurity site
breaks translation symmetry~\cite{senjean2017local}. 
This fact, which is the price to pay for
achieving an exact embedding, explains why 
$\overline{e}_{\rm c}^{\rm bath}(\mathbf{n})$ should in principle depend
not only on the impurity site occupation but {\it also} on the bath site
ones [see Eq.~(\ref{eq:emb_pot_uniform})]. 
For simplicity, the latter dependence will be neglected in the following, 
\begin{eqnarray}\label{eq:iDFA}
\overline{e}_{\rm c}^{\rm bath}(\mathbf{n}) \rightarrow
\overline{e}_{\rm c}^{\rm bath}(n_0),
\end{eqnarray}
or, equivalently [see Eq.~(\ref{eq:local_ecbath_exp})],
\begin{eqnarray}
E_{\rm c}^{\rm imp}(\mathbf{n})\rightarrow E_{\rm c}^{\rm imp}(n_0),
\end{eqnarray}
thus leaving for future work the investigation of
bath-occupations-dependent density-functional approximations.

\subsection{Local density-functional approximations based on Bethe
ansatz solutions}\label{subsec:DFA}

So far, only one density-functional approximation to the impurity correlation
energy [referred to as \textit{two-level impurity 
local density approximation} (2L-ILDA) in Ref.~\cite{senjean2017local}]
has been proposed. It is based on the asymmetric Hubbard dimer and
provides essentially an approximate density-functional embedding
potential that is set to zero in the bath. Since 2L-ILDA does not model
the correlation energy of the bath, it cannot be used straightforwardly
for calculating per-site energies and double occupations. We propose in
the following to use Bethe ansatz solutions to (fully- or
impurity-interacting) infinite systems in order to design local density
approximations to the per-site bath correlation energy.   

\subsubsection{Approximation to $e_{\rm c}(n)$}

Regarding the fully-interacting Hubbard model, the
BALDA~\cite{lima2002density,lima2003density,capelle2003density}
(which is exact in the thermodynamic limit
when $U = 0$, $U \rightarrow +\infty$,
and for all $U$ values when $n=1$)
will be used for modeling $e_{\rm c}(n)$. 
The correlation energy within BALDA reads
\begin{eqnarray}\label{eq:ec_balda}
e_{\rm c}^{\rm BA}(U,t,n)&  =  & e^{\rm BA}(U,t,n) -  e^{\rm BA}(U=0,t,n) - \dfrac{U}{4}n^2, \nonumber \\
\end{eqnarray}
where the $U$- and $t$-dependence of the per-site correlation energy
will be dropped for convenience, and the per-site energy is given by
\begin{eqnarray}\label{eq:persite_balda}
e^{\rm BA}(n \leq 1) = \dfrac{-2t \beta (U/t)}{\pi} \sin \left( \dfrac{\pi n}{\beta (U/t)} \right),
\end{eqnarray}
and
\begin{eqnarray}\label{eq:balda_n_greater_1}
e^{\rm BA}(n \geq 1) & = & e^{\rm BA}(2 - n) + U(n - 1).
\end{eqnarray}
The $U/t$-dependent function $\beta (U/t)$ is determined by solving
\begin{equation}
 \dfrac{-2 \beta (U/t)}{\pi} \sin \left( \dfrac{\pi}{\beta (U/t)} \right) 
 =  -4 \int_0^\infty \frac{dx}{x} 
 \dfrac{J_0(x)J_1(x)}{1 + \exp\left(\frac{U}{2t}x\right)},
\end{equation}
where $J_0$ and $J_1$ are zero and first order Bessel functions.
Although this functional has been proved to give accurate energy and 
density profiles, we show here that it depicts a wrong behaviour 
around $U = 0$ away from the half-filled case, which appears to be important
for the calculation of the double occupation in Eq.~(\ref{eq:dblocc_KS}).
Indeed,
since $\beta(0)=2$ and 
$\partial \beta(U/t) / \partial U |_{U = 0} = -\pi / (8t)$ it comes
\begin{eqnarray}\label{eq:ec_balda_U_zero}
\left.\dfrac{\partial e^{\rm BA}_{\rm c}(n)}{\partial U}\right|_{U = 0}
& = & \dfrac{1}{4}\left[\sin \left( \dfrac{\pi n}{2} \right)-n^2\right]
- \dfrac{n\pi}{8}\cos \left( \dfrac{\pi n}{2}\right),\nonumber\\
\end{eqnarray}
and, consequently, for $n\leq 1$, 
\begin{eqnarray}\label{eq:vc_balda_U_zero}
\dfrac{\partial}{\partial n}\left.\dfrac{\partial e^{\rm BA}_{\rm c}(n)}{\partial U}\right|_{U = 0}
& = & \dfrac{n\pi^2}{16}\sin \left( \dfrac{\pi n}{2} \right)-\dfrac{n}{2}.
\end{eqnarray}
As readily seen from Eqs.~(\ref{eq:ec_balda_U_zero}) and
(\ref{eq:vc_balda_U_zero}), away from half-filling, 
both BALDA correlation
energy and potential
will vary linearly with $U$ in the weakly-correlated regime, which is of course unphysical.
This observation will be important when discussing the performance of
BALDA-based functionals for the calculation of per-site energies and
double occupations in SOET, as well as for the analysis of so-called density-driven
errors (see Sec.~\ref{sec:results}).

Note that, in the thermodynamic limit,
 the correlation potential
$\partial e_{\rm c}(n)/\partial n$ exhibits a
discontinuity at half-filling ($n=1$) so that the exact fundamental gap
can be reproduced in KS-SOFT~\cite{lima2002density}.
The BALDA can model such a discontinuity by
construction, as it can actually be seen from Eq.~(\ref{eq:vc_balda_U_zero}) when $n=1$, 
but this leads to convergence problems around the Mott transition 
phase or in the
Coulomb blockade regime. Solutions have been proposed
using finite temperature~\cite{xianlong2012lattice} or {\it ad-hoc} parameters~\cite{kurth2010dynamical,karlsson2011time,ying2014solving}.
On the other hand, in exact SOET, the complementary
per-site bath correlation
potential $\partial \overline{e}_{\rm c}^{\rm
bath}(\mathbf{n})/\partial n_i$ is {\it not} expected to be
discontinuous neither on the impurity site, where the two-electron
repulsion is treated explicitly, nor in the bath where the standard correlation
potential already contains the discontinuity 
[see Eq.~(\ref{eq:emb_pot_uniform})].
This can be easily shown in the atomic limit (see
Appendix~\ref{sec:derivative discontinuity}).

\subsubsection{Approximations to $E_{\rm c}^{\rm imp}(n_0)$}

\begin{figure}
\begin{center}
\includegraphics[scale=0.7]{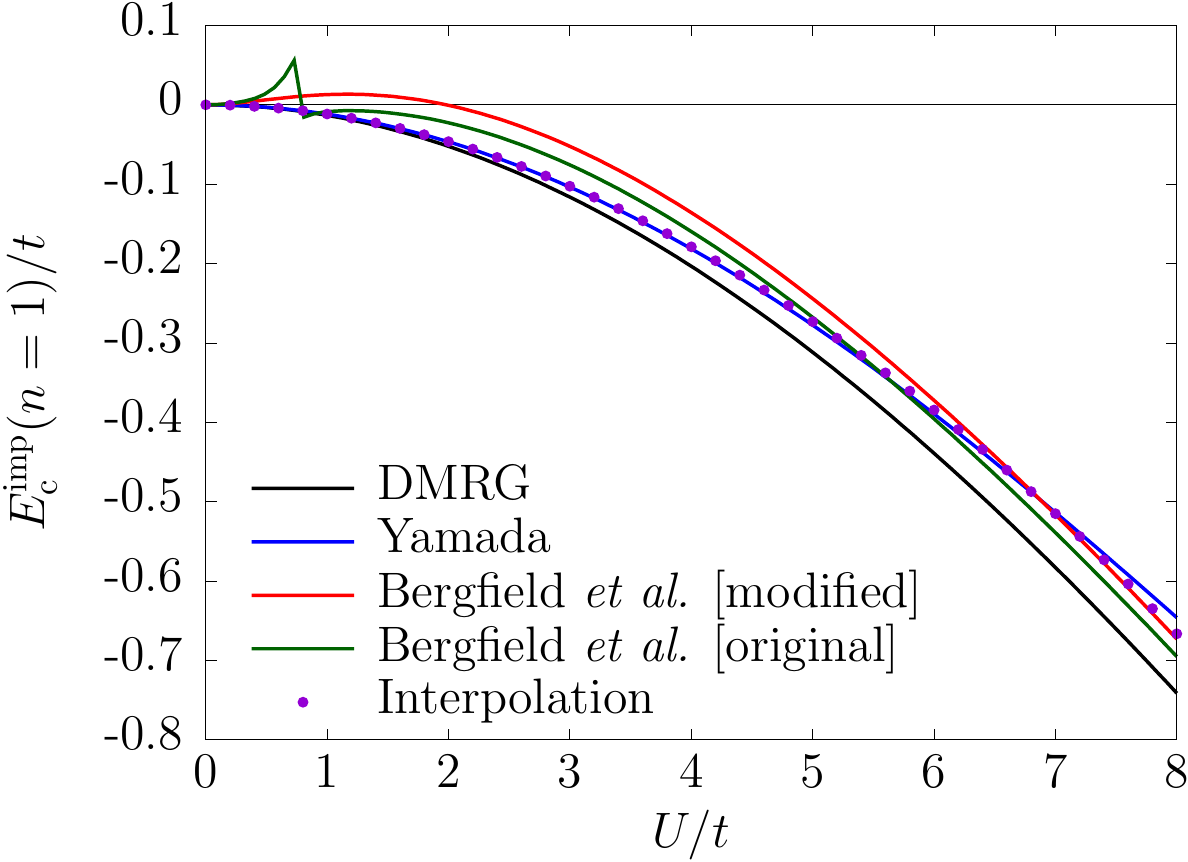}
\caption{Correlation energy of the embedded impurity for the half-filled
 32-site one-dimensional
Hubbard 
model. Various approximations are tested for
$\Gamma=t$ [see Sec.~\ref{subsec:Hubbard_SIAM}]: Eq.~(\ref{eq:anderson_yamada}) 
[blue curve], Eq.~(\ref{eq:anderson_energy}) combined with 
Eq.~(\ref{eq:sigma_Burke_exp}) [green curve], and 
Eq.~(\ref{eq:anderson_energy}) combined with
Eq.~(\ref{eq:sigma_modified}) [red curve]. The interpolation
SIAM-BALDA[$n$=1] relies on [Eq.~(\ref{eq:inter_siam_func}) combined with 
Eq.~(\ref{eq:sigma_modified})] is shown with points in purple. 
The accurate DMRG result [see
Ref.~\cite{senjean2017local} for further details about the accurate calculation
of correlation energies] is shown in black for
comparison. 
}
\label{fig:Ecimp_alphasigma}
\end{center}
\end{figure}

Turning to density-functional approximations for $\overline{e}_{\rm
c}^{\rm bath}(n_0)$ or, equivalently, $
E_{\rm c}^{\rm imp}(n_0)$, 
the simplest one 
[referred to as impurity-BALDA (iBALDA) in the
following]
consists in modeling the correlation energy
of the impurity-interacting system with the BALDA:
\begin{eqnarray}\label{eq:iBALDA_Ecimp}
E_{\rm c}^{\rm imp}(n_0) \xrightarrow{\rm iBALDA} e_{\rm c}^{\rm BA}(n_0).
\end{eqnarray}
In other
words, 
the iBALDA
neglects the contribution of the bath to the total per-site correlation
energy [see Eq.~(\ref{eq:local_ecbath_exp})], 
\begin{eqnarray}\label{eq:ibalda_approx}
\overline{e}_{\rm c}^{\rm bath}(n_0) \xrightarrow{\text{iBALDA}} 0.
\end{eqnarray}
Despite its apparent simplicity, 
this approximation will show to be very accurate away from half-filling,
but it overestimates the correlation energy of the impurity otherwise.
This will be discussed further in Sec.~\ref{sec:results}. 
Improvement can be considered either 
by increasing the number of impurities~\cite{fromager2015exact} 
(and still use the iBALDA),
in analogy with the
DMET~\cite{knizia2012density},
or by 
developing more accurate 
approximations to $\overline{e}_{\rm c}^{\rm bath}(n_0)$
while keeping a single impurity site. The latter
option is of course preferable in terms of computational cost. It can be
implemented, in the {\it half-filled} case, by exploiting the already
mentioned (see Sec.~\ref{subsec:exact_expressions}) analogy between the
auxiliary impurity-interacting system and the    
symmetric SIAM. Using the
latter for extracting an
approximate $E_{\rm c}^{\rm imp}({n}_0)$ functional gives, when combined with
the BALDA, an approximation that will be referred to as the SIAM-BALDA[$n$=1]
in the following,  
\begingroup
\begin{eqnarray}\label{eq:siam-balda}
\overline{e}_{\rm c}^{\rm bath}(n_0=1)
&\xrightarrow[{\rm BALDA} \lbrack n=1 \rbrack] {\rm SIAM}&
e^{\rm BA}_{\rm c}(n_0=1)
\nonumber\\
&&-E^{\rm SIAM}_{\rm c}(U,\Gamma,n_0=1),
\end{eqnarray}
\endgroup
where $\Gamma$ is the impurity level width parameter of the
SIAM~\cite{CRP16_Georges_impurities}. For clarity, we postpone to
Sec.~\ref{subsec:Hubbard_SIAM} the discussion on the choice of $\Gamma$ in the context of SOET.  
The correlation energy in the symmetric SIAM 
can be well described in all
correlation regimes by a simple interpolation 
between the
weakly and the strongly
correlated limits,
\begin{eqnarray}\label{eq:sym_siam_inter}
E^{\rm SIAM}_{\rm c}(U,\Gamma,n=1) 
&=& 
\frac{1}{1+f}
E^{\rm SIAM}_{{\rm c},U/\Gamma\rightarrow 0}(U,\Gamma) 
\nonumber \\ && {}
+ 
\frac{f}{1+f}
E^{\rm SIAM}_{{\rm c}, U/\Gamma\rightarrow\infty}(U,\Gamma,n=1),
\nonumber \\
\label{eq:inter_siam_func}
\end{eqnarray}
where $f=f(U/\Gamma)=e^{U/\Gamma-6.876}$.
Here in the weakly correlated limit,
we use Yamada's perturbative expression through fourth order in
$U/\Gamma$~\cite{yamada1975perturbation},
\begin{equation}
E^{\rm SIAM}_{{\rm c}, U/\Gamma\rightarrow 0}(U,\Gamma) 
 = 
\dfrac{U^2}{\pi\Gamma}
\left[- 0.0369 
 + 0.0008 \left( \dfrac{U}{\pi
\Gamma} \right)^2\right].
\label{eq:anderson_yamada}
\end{equation}
Regarding the strongly correlated limit,
we propose to use a simplified version of the density-functional approximation
developed by Bergfield {\it et al.} which relies on the
BA solution to the strongly correlated
SIAM~\cite{bergfield2012bethe,liu2012accuracy}.
An impurity correlation energy functional is obtained by integrating
(with respect to the density $n$) the parameterized correlation
potential of Eqs.~(15) and (16) in Ref.~\cite{bergfield2012bethe}, which gives:
\begin{widetext}
\begingroup
\begin{eqnarray}
\begin{array}{l}
E^{\rm SIAM}_{{\rm c}, U/\Gamma\rightarrow\infty}(U,\Gamma,n)
= \alpha(U,\Gamma) \dfrac{U}{2}\Big[\mathcal{E}_{\rm c}
(U,\Gamma,n)-\mathcal{E}_{\rm c}(U,\Gamma,0)\Big],
\\
\\
\mathcal{E}_{\rm c}(U,\Gamma,n)=n - \dfrac{n^2}{2} + 
\dfrac{2}{\pi}(1 - n){\rm tan}^{-1}\left[\dfrac{(1 - n)}{\sigma} \right]
 - \dfrac{\sigma}{\pi} \ln \left[ 1 + \left(\dfrac{(1-n)}{\sigma}\right)^2 \right]  
,
\end{array}
\label{eq:anderson_energy}
 \end{eqnarray}
 \endgroup
\end{widetext}
 where $\alpha(U,\Gamma) = U/(U+5.68\Gamma)$ and
\begin{eqnarray}\label{eq:sigma_Burke_exp}
\sigma = 0.811 \dfrac{\Gamma}{U}
-0.39 \left(\dfrac{\Gamma}{U}\right)^2
-0.168 \left(\dfrac{\Gamma}{U}\right)^3.
\end{eqnarray}
Note that, in order to be able to use the interpolation in
Eq.~(\ref{eq:sym_siam_inter}) in any regime of correlation, i.e., for all
values of $U/\Gamma$, we need to reconsider the parameterization in 
Eq.~(\ref{eq:sigma_Burke_exp}). Indeed, when passing through $\sigma=0$ or,
equivalently, when $U/\Gamma\approx  0.755$, the correlation energy undergoes
a jump because of the $\tan^{-1}$ function, as shown in
Fig.~\ref{fig:Ecimp_alphasigma}. For this reason, we will use in the
SIAM-BALDA[$n$=1] approximation the simpler (and still reasonably
accurate) expression,    
\begin{eqnarray}\label{eq:sigma_modified}
\sigma \xrightarrow[{\rm BALDA}\lbrack n = 1 \rbrack]{{\rm SIAM}} \dfrac{8\Gamma}{\pi^2 U},
\end{eqnarray}
which originates from the BA solution as $U\rightarrow+\infty$~\cite{bergfield2012bethe}.
Note that, with this choice, $E^{\rm SIAM}_{{\rm c},
U/\Gamma\rightarrow\infty}(U,\Gamma,n=1)$ becomes positive for smaller
$U/\Gamma$ values, which is of course unphysical. This artefact is actually removed by the
interpolation in Eq.~(\ref{eq:sym_siam_inter}), as shown in
Fig.~\ref{fig:Ecimp_alphasigma}. The numerical value $6.876$
in the interpolation function $f$ simply corresponds to the crossing
point between Yamada's [Eq.~(\ref{eq:anderson_yamada})] and modified
Bergfield's [Eqs.~(\ref{eq:anderson_energy}) and
(\ref{eq:sigma_modified})] approximate correlation energies. 

\subsection{Connecting SOET to the SIAM}\label{subsec:Hubbard_SIAM}

In order to use density-functional approximations based on the SIAM in
the context of SOET we need to relate the impurity level width parameter $\Gamma$
of the SIAM to the parameters of the (original) Hubbard problem $t$ and
$U$, and,
possibly, the density.
This is the purpose of this section. Let us start with the expression
for the Hamiltonian of
the symmetric SIAM written in (discretized) real
space,
\begin{eqnarray}
\label{eq:siam_ham_real_space}
\hat{H}^{\rm SIAM}&=&
-t\sum_\sigma\sum^{L}_{i=1}
\left(
\hat{c}^\dagger_{i\sigma}\hat{c}_{i+1\sigma}+
\mathrm{H.c.}
\right)
\nonumber
\\
&&+V\sum_\sigma\left(\hat{c}^\dagger_{1\sigma}\hat{d}_{\sigma}+
\hat{c}^\dagger_{L\sigma}\hat{d}_{\sigma}
+\mathrm{H.c.}
\right)
\nonumber\\
&& +U\hat{n}_{d\uparrow}\hat{n}_{d\downarrow}- \dfrac{U}{2}\hat{n}_{d}, 
\end{eqnarray}
where we denote $\hat{d}_{\sigma}=\hat{c}_{0\sigma}$,
$\hat{n}_{d\sigma}=\hat{d}^\dagger_{\sigma}\hat{d}_{\sigma}$, and
$\hat{n}_{d}=\sum_\sigma\hat{n}_{d\sigma}$. 
We assume the periodic boundary condition in the bath
$\hat{c}_{L+1\sigma}=\hat{c}_{1\sigma}$.
As discussed in Sec.~\ref{subsec:exact_expressions},
in the particular case of a
half-filled $(L+1)$-site Hubbard problem, the exact auxiliary impurity-interacting Hamiltonian of SOET 
is essentially the one in Eq.~(\ref{eq:siam_ham_real_space}) if
\begin{eqnarray}\label{eq:V_eq_minus_t}
V=-t. 
\end{eqnarray}  
Note that, in principle, we should remove the coupling term between the two neighbors
($i=1$ and $i=L$) of the impurity site. The latter point is ignored in the
following for simplicity. 
Using 
the representation of bath
creation operators in $k$-space,
\begin{eqnarray}
\hat{c}^\dagger_{i\sigma}=\frac{1}{\sqrt{L}}\sum_ke^{-{\rm
i}ki}\hat{c}^\dagger_{k\sigma},
\end{eqnarray}
with 
\begin{eqnarray}
k=\frac{2\pi}{L} m \quad
\left(
m=-\frac{L}{2}+1,\ldots,\frac{L}{2}
\right),
\label{eq:quantized_k_values}
\end{eqnarray}
we recover from Eq.~(\ref{eq:siam_ham_real_space}) the usual SIAM
Hamiltonian expression,   
\begin{eqnarray}\label{eq:siam_ham_k_space}
\hat{H}^{\rm SIAM}&=&
\sum_{k\sigma}\varepsilon_k\hat{c}^\dagger_{k\sigma}\hat{c}_{k\sigma}
+U\hat{n}_{d\uparrow}\hat{n}_{d\downarrow}-\dfrac{U}{2}\hat{n}_{d}
\nonumber\\
&&+\sum_{k\sigma}\Big(V(k)\hat{c}^\dagger_{k\sigma}\hat{d}_{\sigma}+\mathrm{H.c.}
 \Big)
,
\end{eqnarray} 
where $\varepsilon_k=-2t \cos k$ and 
\begin{eqnarray}
\label{eq:V_k_siam}
V(k)=
\frac{2V}{\sqrt{L}} \,
e^{-{\rm i}k/2}\, \cos (k/2)
\end{eqnarray}
is the $k$-dependent coupling term
between the bath and the impurity. 
The correlation energy of the impurity is then determined
from the frequency-dependent hybridization
function~\cite{CRP16_Georges_impurities} 
\begin{eqnarray}
\Gamma(\omega)=\pi\sum_k\vert
V(k)\vert^2\delta(\omega-\varepsilon_k),
\end{eqnarray}
which, according to Eqs.~(\ref{eq:quantized_k_values}) and (\ref{eq:V_k_siam}), can be simplified as follows in the
thermodynamic limit ($L\rightarrow+\infty$),
\begin{eqnarray}\label{eq:Gamma_exp_omega}
\Gamma(\omega)&=&
\dfrac{L}{2}\int^\pi_{-\pi} d k \vert
V(k)\vert^2\delta(\omega-\varepsilon_k)
\nonumber\\
&=&
4V^2\int^\pi_{0} d k\, \cos^2(k/2)\delta(\omega-\varepsilon_k)
\nonumber\\
&=&
\dfrac{V^2}{t^2}
\int^{2t}_{-2t} d\varepsilon
\dfrac{t-\dfrac{\varepsilon}{2}}{\sqrt{1-\dfrac{\varepsilon^2}{4t^2}}}
\delta(\omega-\varepsilon).
\end{eqnarray}
The (frequency-independent) impurity level width parameter $\Gamma$ of
the SIAM is usually defined as the value of the
hybridization function at the Fermi
level $\varepsilon_F=-2t\,\cos  k_F$, 
\begin{eqnarray}
\label{eq:Gamma_of_Fermi}
\Gamma=\Gamma(\varepsilon_F)
=
\dfrac{V^2}{t^2}
\dfrac{t-\dfrac{\varepsilon_F}{2}}{\sqrt{1-\dfrac{\varepsilon_F^2}{4t^2}}}
.
\end{eqnarray}
By using Eq.~(\ref{eq:V_eq_minus_t}) and the relation between the uniform
density $n=N/L$ in the
bath and $k_F$,  
\begin{eqnarray}\label{eq:dens_k_int}
N = 2 \cdot \frac{L}{2\pi} \int_{-k_F}^{k_F} dk,
\end{eqnarray}
or, equivalently,
\begin{eqnarray}\label{eq:k_F_dens}
n=\frac{2}{\pi}k_F,
\end{eqnarray}
we finally obtain a $t$-dependent density-functional impurity
level width which connects the SIAM to the original Hubbard problem to
be solved in SOET:  
\begin{eqnarray}\label{eq:Gamma_fun}
\Gamma=\Gamma(t,n)&=&
t\left(\frac{1+\cos (\pi n/2)}{\sin (\pi n/2)}\right).
\end{eqnarray}
Note that the latter expression is valid when
$0\leq n \leq 1$. In the range $1\leq n\leq 2$, we should use the
hole-particle symmetry relation $\Gamma(t,n)=\Gamma(t,2-n)$.

As readily seen from Eq.~(\ref{eq:Gamma_fun}), we obtain $\Gamma=t$ for
the half-filled Hubbard problem ($n=1$). This is the reason why, in
Sec.~\ref{sec:results}, per-site correlation energies have been
computed for the bath at the SIAM-BALDA[$n$=1] level of approximation (see
Eq.~(\ref{eq:siam-balda})) with $\Gamma$ set to $t$. 
The deviation from half-filling in the original Hubbard system will be
interpreted, in the SIAM, as a rescaling of $\Gamma$. In
the low-density regime we have 
$\Gamma(t,n)\approx 4t/(\pi n) \gg \Gamma(t,n=1)$,
thus leading  
to weaker correlation effects on the embedded impurity site, in
comparison to the half-filled case. Consequently, we might expect the simple
combination of Yamada's perturbation expansion in $U/\Gamma$ [see
Eq.~(\ref{eq:anderson_yamada})] with Eq.~(\ref{eq:Gamma_fun})
to provide a reasonable approximation to 
the correlation energy of the impurity, even when entering the
strong correlation regime (this point will be further discussed in the
following).
The latter approximation combined
with BALDA, for the calculation of the per-site correlation energy of the
bath, will
be
referred to as SIAM-BALDA (without the suffix 
[$n$=1]) in the following.
It can be summarized as follows,
\begin{equation}
\label{eq:siam_balda_gammafctn}
\overline{e}_{\rm c}^{\rm bath}(n_0)
\xrightarrow[{\rm BALDA}] {{\rm SIAM}}
e^{\rm BA}_{\rm c}(n_0)
-E^{\rm SIAM}_{{\rm c}, U/\Gamma \rightarrow 0}(U,\Gamma(t,n_0)).
\end{equation}
In contrast to its [$n$=1] analog, SIAM-BALDA is applicable to {\it any} density regime.
Note that, for $n=1$, SIAM-BALDA and SIAM-BALDA[$n$=1] will {\it not}
give exactly the same result. The difference will become substantial in the
strongly correlated regime where, by construction, the latter
approximation will be 
more accurate than the former. 

\section{Summary of the various density-functional approximations and computational details}\label{sec:Computational details}

In order to perform practical SOET calculations we 
must solve the self-consistent impurity problem in
Eq.~(\ref{eq:self-consistent-SOET}) where, as readily seen from
Eq.~(\ref{eq:emb_pot_uniform}), density-functional approximations to the total and bath per-site
correlation energies, i.e., $e_{\rm c}(n)$ and $\overline{e}_{\rm c}^{\rm
bath}({\bf n})$, are needed. In our calculations, the original 
one-dimensional
uniform Hubbard system
will consist of 32 sites. The embedded impurity wavefunction 
[which is the solution to the self-consistent Eq.~(\ref{eq:self-consistent-SOET})] has
been computed accurately (either fully self-consistently or, for
analysis purposes, by inserting
the exact uniform density into the complementary Hxc bath potential) by applying the DMRG method
\cite{white1992density,white1993density,verstraete2008matrix,
schollwock2011density,naokicode} to the density-functional
impurity-interacting Hamiltonian in Eq.~(\ref{eq:self-consistent-SOET}).
The maximum number of renormalized states (or virtual bond
dimension) was set to $m = 500$. Turning to the functionals, BALDA
[see Eq.~(\ref{eq:ec_balda})] has
been used for the total per-site correlation energy in both SOET and
conventional (KS) SOFT calculations. Regarding the complementary 
correlation energy of
the bath, various approximations have been considered: iBALDA
[Eq.~(\ref{eq:ibalda_approx})], the interpolation-based
SIAM-BALDA[$n$=1] for calculations at half-filling
[Eqs.~(\ref{eq:siam-balda}), (\ref{eq:sym_siam_inter}), and
(\ref{eq:sigma_modified}) with $\Gamma=t$], and SIAM-BALDA
[Eq.~(\ref{eq:siam_balda_gammafctn})]. 
The expressions for the various density-functional
approximations are summarized 
in Table~\ref{tab}.
Finally,
Eqs.~(\ref{eq:per-site-energy}) and (\ref{eq:doubleocc_soet}) have been
implemented for the calculation of total per-site energies and
double occupations, respectively. The same quantities have been computed
in SOFT by implementing Eqs.~(\ref{eq:per-site-energy_KS}) and 
(\ref{eq:dblocc_KS}). Standard DMRG calculations, where 
the DMRG method is applied to the physical fully-interacting uniform Hubbard
Hamiltonian in Eq.~(\ref{eq:Hubb_hamiltonian}) [the potential ${\bf v}$ is set
to zero in this case], are used as reference in the following. They will
simply be referred
to as ``DMRG'' in the rest of this work.  
The SOET calculations, where DMRG is applied to the embedded-impurity system, will be referred to by the name of the Hxc bath
functional that is used (iBALDA, SIAM-BALDA[$n$=1] or SIAM-BALDA). 

\begin{table*}
\begin{center}
\begin{tabularx}{1\textwidth}{c *{2}{Y}}
\toprule
\hline\hline
 \multicolumn{1}{c}{SOET method}
 & \multicolumn{1}{c}{Density-functional approximation used for $\overline{E}_{\rm Hxc}^{\rm bath}(\mathbf{n})$} 
 & \multicolumn{1}{c}{Correlation density-functional approximations} \\ \\
\hline
 \\
iBALDA & $\displaystyle \sum_{i\neq 0} \left(\dfrac{U}{4}n_i^2 + e_{\rm c}^{\rm BA}(n_i)\right)$
& Eqs.~(\ref{eq:ec_balda})--(\ref{eq:balda_n_greater_1})\\ \\
SIAM-BALDA[$n=1$] & 
$\displaystyle \sum_{i} \left(\dfrac{U}{4} + e_{\rm c}^{\rm BA}(1) \right) 
-\dfrac{U}{4}-E^{\rm SIAM}_{\rm c}(U,\Gamma=t,1)$
& Eqs.~(\ref{eq:ec_balda})--(\ref{eq:balda_n_greater_1}),
(\ref{eq:sym_siam_inter})--(\ref{eq:anderson_energy}), and (\ref{eq:sigma_modified})  
\\ \\
SIAM-BALDA & $\displaystyle \sum_{i} \left(\dfrac{U}{4}n_i^2 + e_{\rm c}^{\rm BA}(n_i) \right) 
-\dfrac{U}{4}n_0^2-E^{\rm SIAM}_{{\rm c},U/\Gamma\rightarrow
0}\left(U,\Gamma(t,n_0)\right)$
&Eqs.~(\ref{eq:ec_balda})--(\ref{eq:balda_n_greater_1}),
(\ref{eq:anderson_yamada}), and (\ref{eq:Gamma_fun})\\ \\
\hline
\hline
\end{tabularx}
\caption{Summary of the density-functional approximations used for 
$\overline{E}_{\rm Hxc}^{\rm bath}(\mathbf{n})$
in the practical SOET
calculations (see Eqs.~(\ref{eq:variational_energy}) and
(\ref{eq:self-consistent-SOET}))
presented in this work. The corresponding
approximate bath Hxc potentials on site $i$ are simply obtained by differentiation with respect to 
$n_i$. In (half-filled) SIAM-BALDA[$n=1$]
calculations, we used the potential $\partial \overline{E}_{\rm Hxc}^{\rm
bath}(\mathbf{n})/\partial n_i=(1-\delta_{i0})U/2$ which is exact for
{\it finite-size} half-filled uniform rings.}
\label{tab}
\end{center}
\end{table*}

\section{Results and discussion}\label{sec:results}

\begin{figure}
\begin{center}
\includegraphics[scale=0.6]{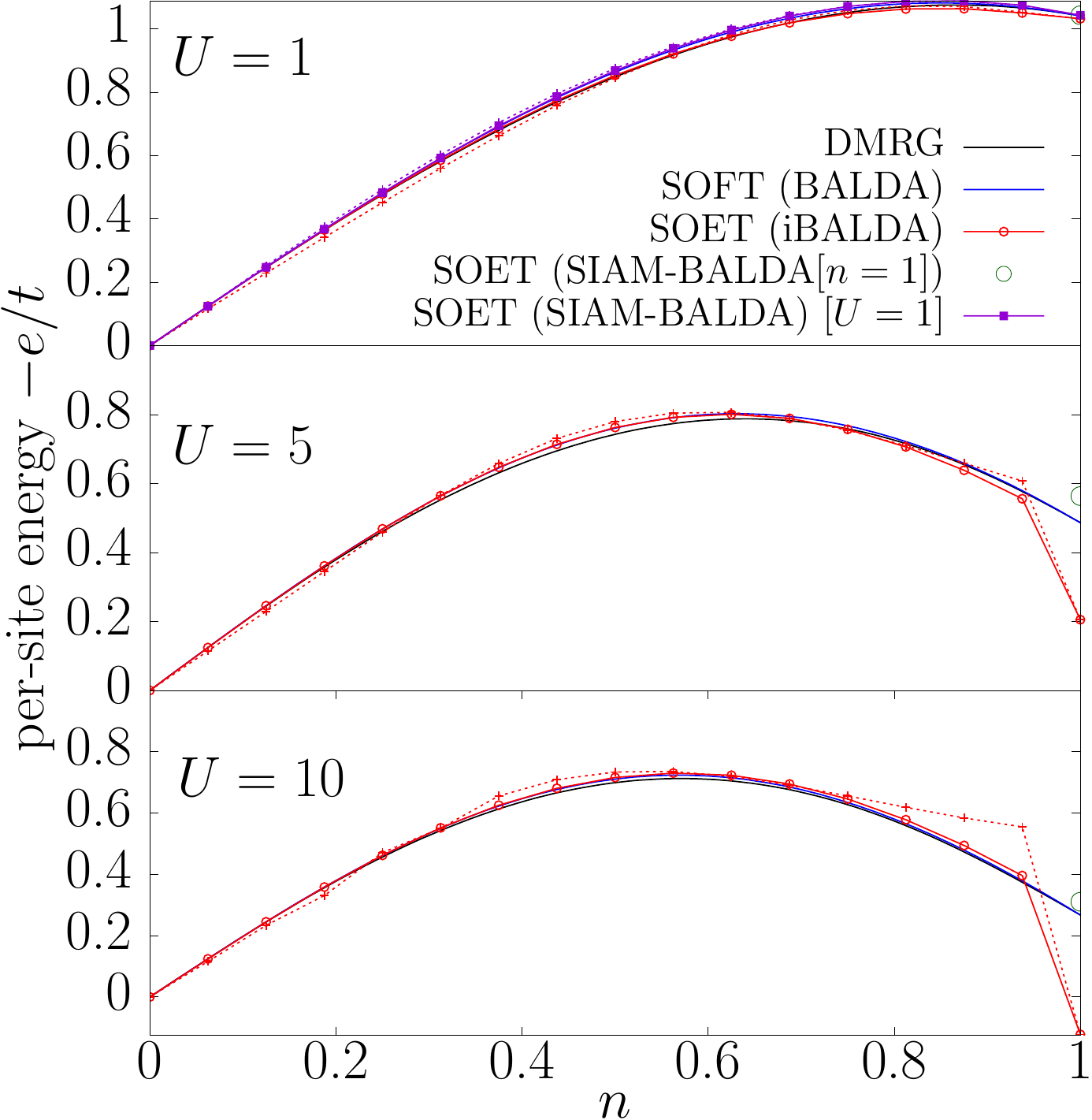}
\end{center}
\caption{Total per-site energies ($-e/t$) plotted as a function of the exact density $n = N/L$ for $U=1, 5$
and $10$, $t=1$ and $L=32$.
Results obtained with SIAM-BALDA[$n$=1], which is only defined at
half-filling, are shown with green open circles. Results obtained with
SIAM-BALDA are only shown for $U=1$ (see text for further details). 
iBALDA as well as SIAM-BALDA [$U$=1] energies obtained with self-consistently converged
densities are plotted with dashed lines. Comparison is made with SOFT
(BALDA) and DMRG.
}
\label{fig:persite-energy}
\end{figure}

\begin{figure}
\begin{center}
\includegraphics[scale=0.57]{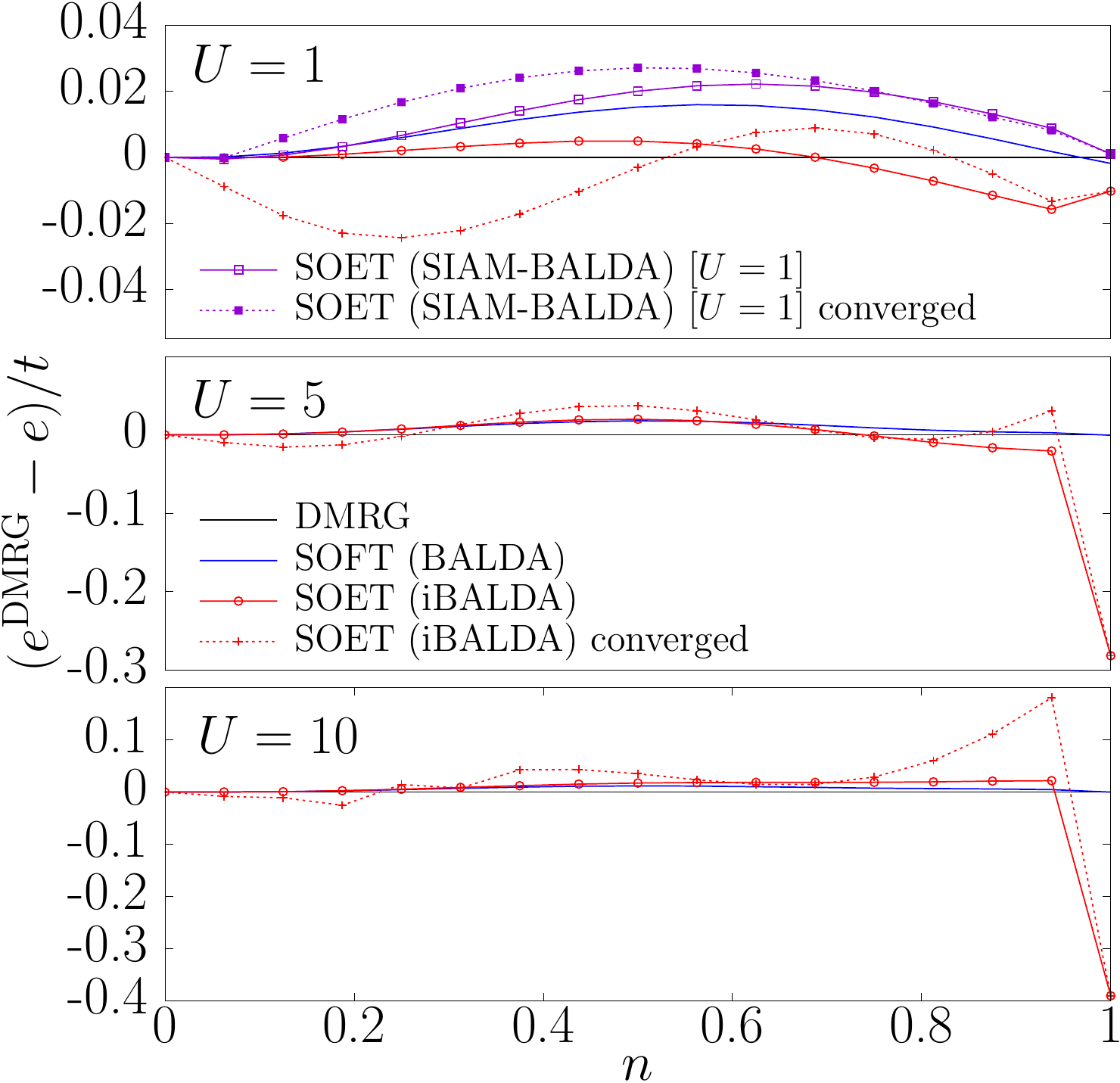}
\end{center}
\caption{Similar to Fig.~\ref{fig:persite-energy} but deviations from 
the DMRG result are plotted (instead of total per-site energies) for ease of
comparison. Dashed lines are used for SOET results obtained with
self-consistently converged densities. 
}
\label{fig:persite-error}
\end{figure}

\begin{figure}
\begin{center}
\includegraphics[scale=0.7]{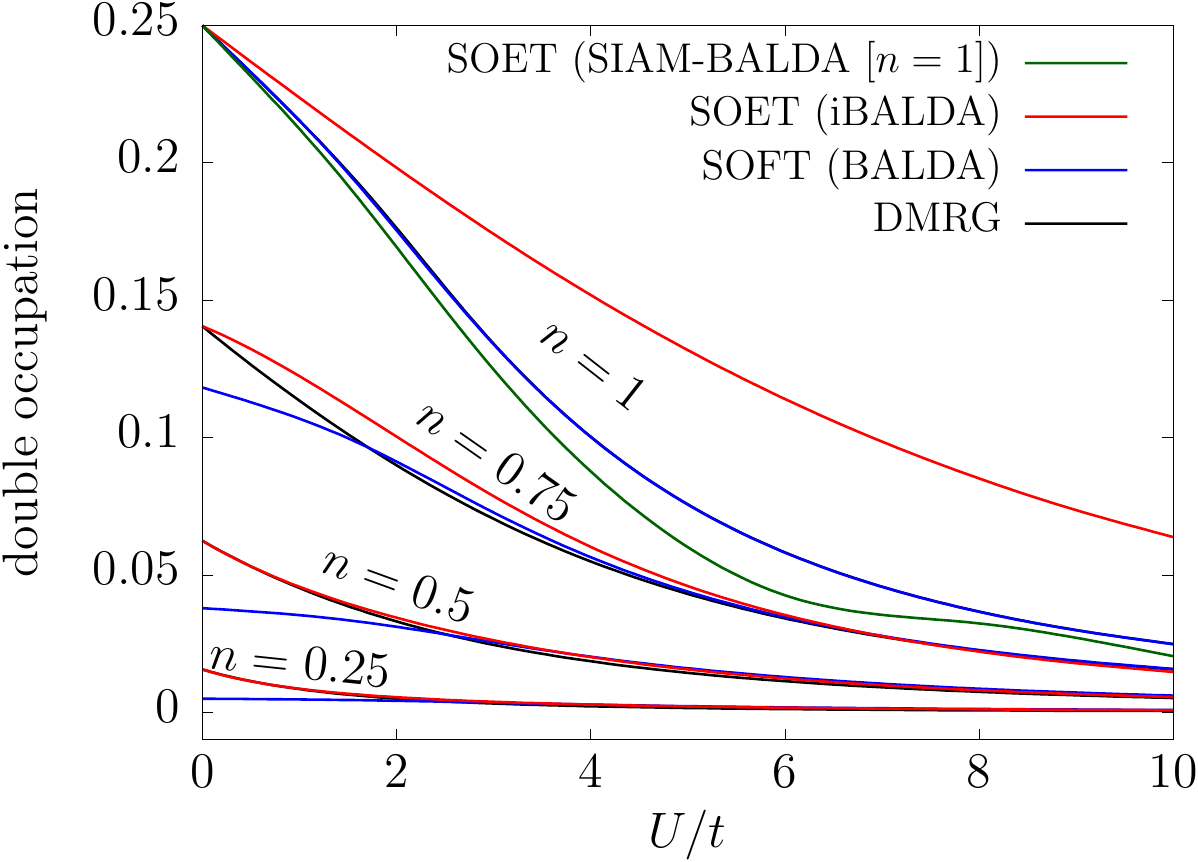}
\end{center}
\caption{Double occupations plotted as a function of $U/t$ for various
exact densities $n = N/L$ with $L=32$. Results obtained with SIAM-BALDA[$n$=1],
which is only defined at half-filling, are plotted for $n= 1$ (green
curve).
iBALDA double occupations obtained with self-consistently converged densities (not shown) are almost indistinguishable from
the ones obtained with the exact densities. 
Note that, for $n = 1$, 
the BALDA curve is on top of the DMRG curve,
as expected.}
\label{fig:dblocc}
\end{figure}

Let us first focus on the performance of iBALDA for the calculation of
total per-site energies [Figs.~\ref{fig:persite-energy} and
\ref{fig:persite-error}] and double occupations [Fig.~\ref{fig:dblocc}].
Even though it performs well away from half-filling for all $U/t$ values, 
the iBALDA underestimates the correlation energy significantly in the strongly correlated regime
when approaching half-filling ($n = 1$),
as shown in the bottom panel of Fig.~\ref{fig:persite-energy}. This is also
reflected in the double occupation (see Fig.~\ref{fig:dblocc}) which,
interestingly, is 
comparable to the one obtained at the 1-site DMET level~\cite{knizia2012density}.
As expected, both per-site energies and double occupations 
are significantly improved 
when applying the SIAM-BALDA[$n$=1] functional
(see 
Figs.~\ref{fig:persite-energy} and \ref{fig:dblocc}).
Let us stress that, in
order to obtain similar results in DMET, one would need to increase the
number of impurity sites~\cite{knizia2012density} while, at the SIAM-BALDA[$n$=1] level of
approximation, we keep on using a single impurity site.

\begin{figure}
\begin{center}
\includegraphics[scale=0.6]{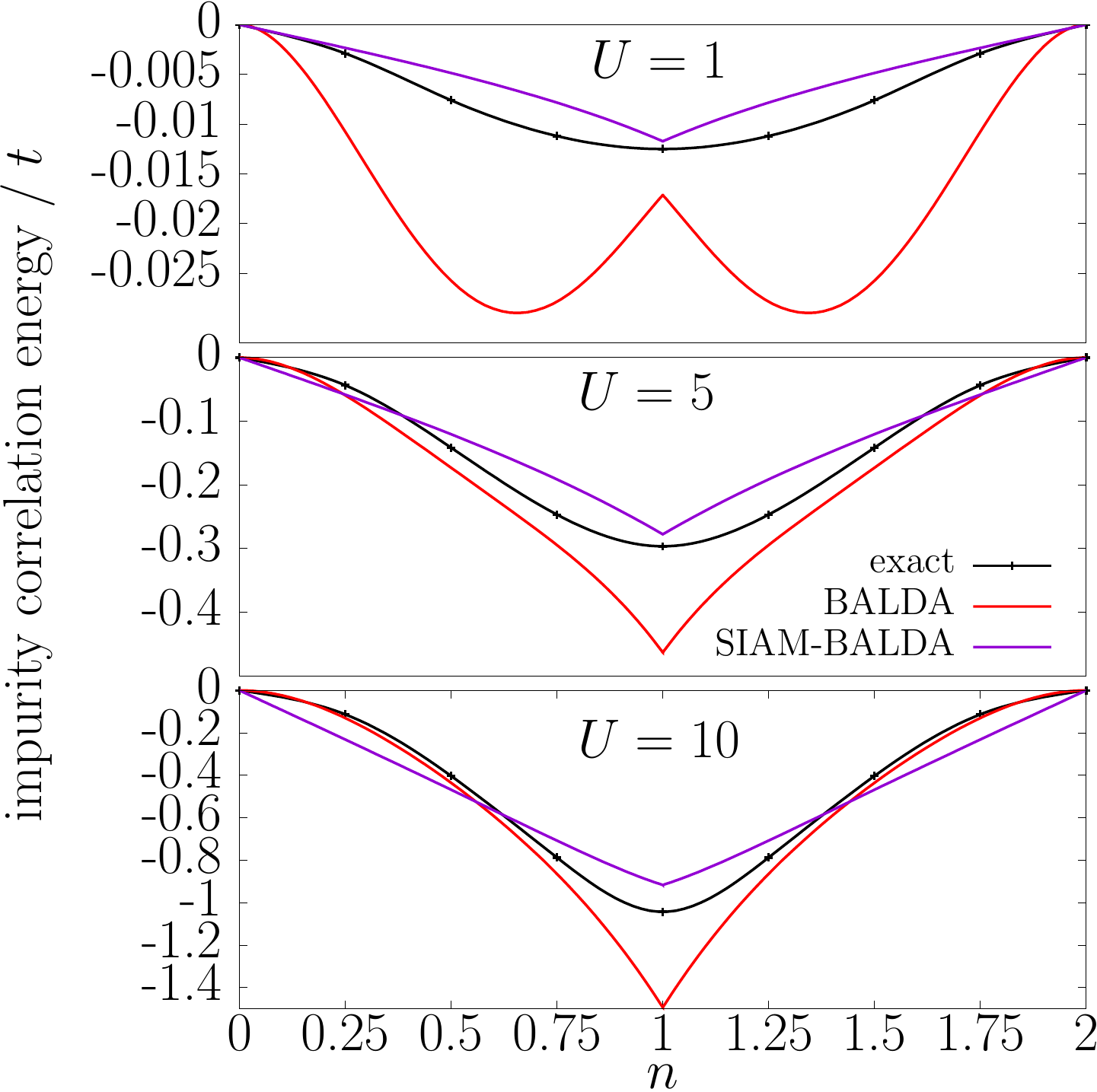}
\caption{Comparing the density-functional impurity correlation
energy used in SIAM-BALDA [see Eqs.~(\ref{eq:anderson_yamada}) and
(\ref{eq:Gamma_fun})] with the exact results of
Ref.~\cite{senjean2017local}. The latter were obtained for a 8-site ring. The (total) BALDA per-site correlation
energy is shown for analysis purposes.}
\label{fig:Ecimp_wrtn_gammafctn}
\end{center}
\end{figure}

Away from half-filling, SOFT (BALDA)
systematically underestimates the double occupation in the
weakly-correlated regime, as shown in
Fig.~\ref{fig:dblocc}. 
This is a direct consequence of the unphysical 
linear behavior in $U$ of $e^{\rm BA}_{\rm c}(n)$ when $U\rightarrow 0$
and $n\neq 1$ [see Eq.~(\ref{eq:ec_balda_U_zero})]. On the other hand, the iBALDA, which uses the ``bare'' double
occupation of the embedded impurity 
[see Eqs.~(\ref{eq:doubleocc_soet}) and (\ref{eq:ibalda_approx})]
recovers the exact double occupation in the weakly correlated limit ($U = 0$) and
gives
relatively accurate results otherwise.
Note that exact densities have been used so far. 
By solving
Eq.~(\ref{eq:self-consistent-SOET})
self-consistently within the iBALDA,
we introduce
density-driven errors 
in the per-site energy (see the dashed lines in Fig.~\ref{fig:persite-error}). 
In the weakly correlated regime, they are 
clearly related to the unphysical linearity in $U$ of the BALDA
correlation potential 
[see Eq.~(\ref{eq:vc_balda_U_zero})]. 
Note that, at the iBALDA level of approximation, the double
occupations obtained self-consistently (not shown) are essentially on
top of the ones obtained with the exact densities, simply because the density-functional contribution is neglected.
Let us finally mention that, in the vicinity of the half-filled strongly-correlated regime, the iBALDA
per-site energy deteriorates if self-consistently converged densities
are used (see Fig.~\ref{fig:persite-energy}). As shown in 
Fig.~\ref{fig:Ecimp_wrtn_gammafctn} (see the $U=5$ and $U=10$ panels), the BALDA per-site correlation energy differs substantially from the exact impurity
correlation one, especially around $n=1$, thus making the iBALDA approximation
irrelevant in this regime of correlation and density. 
SIAM-BALDA gives, on the other hand,
a far more accurate description of the impurity correlation energy than BALDA around half-filling, as
illustrated in Fig.~\ref{fig:Ecimp_wrtn_gammafctn}. 
Note that, at low
density, SIAM-BALDA does not give very accurate impurity
correlation energies for large $U$ values, thus somehow invalidating our
assumption (see Sec.~\ref{subsec:Hubbard_SIAM}) that, at low densities, combining Yamada's expansion in $U/\Gamma$ of the SIAM
correlation energy with the density-dependent impurity level
width parameter $\Gamma(t,n)$ of Eq.~(\ref{eq:Gamma_fun}) would be
sufficient. A better approximation is clearly needed
in this regime of correlation and density.

Let us now briefly discuss the performance of SIAM-BALDA. 
We only show results obtained with the relatively small $U=1$ value for which, at
half-filling, Yamada's
perturbation expansion of the SIAM correlation energy is accurate.   
Although, as discussed previously, SIAM-BALDA 
provides an overall better description of the density-functional impurity
correlation energy than BALDA, even in stronger correlation regimes,
its combination with BALDA (see Eq.~(\ref{eq:siam_balda_gammafctn})) for the
calculation of per-site energies and double occupations will not
necessarily provide good results as $U/t$ increases. Indeed, as readily seen in
Eqs.~(\ref{eq:doubleocc_soet}) and (\ref{eq:per-site-energy}),
it is in principle important to reproduce the proper 
dependence in $t$ and $U$ of the complementary per-site bath correlation functional.
Moreover, the density dependence of the SIAM-BALDA impurity correlation
energy is far from satisfactory (see
Fig.~\ref{fig:Ecimp_wrtn_gammafctn}), which may lead to substantial density-driven
errors. The performance of SIAM-BALDA in stronger 
correlation regimes will be discussed further in a forthcoming paper.     
Obviously, the same criticism would apply to the interpolation in Eq.~(\ref{eq:sym_siam_inter}) 
where the density-dependent $\Gamma(t,n)$ of Eq.~(\ref{eq:Gamma_fun})
could be used for {\it any} density $n$ (not just $n=1$). Such
a choice would be pragmatic since the interpolation is only justified in
the half-filled case. A density-dependent generalization of the interpolation formula  
would be needed in order to obtain a density-functional approximation that is
applicable to any density and correlation regime.
Let us finally point out that the use of a density-dependent impurity
level width parameter $\Gamma(t,n)$ in the strongly correlated limit of
the SIAM correlation functional
$E^{\rm SIAM}_{{\rm c}, U/\Gamma\rightarrow\infty}(U,\Gamma,n)$ might enhance the density
dependence of the embedding potential. It is unclear how physical (or
unphysical) this can be. This should obviously be analyzed in detail. We
keep such an analysis for future work. 

\begin{figure}
\begin{center}
\includegraphics[scale=0.6]{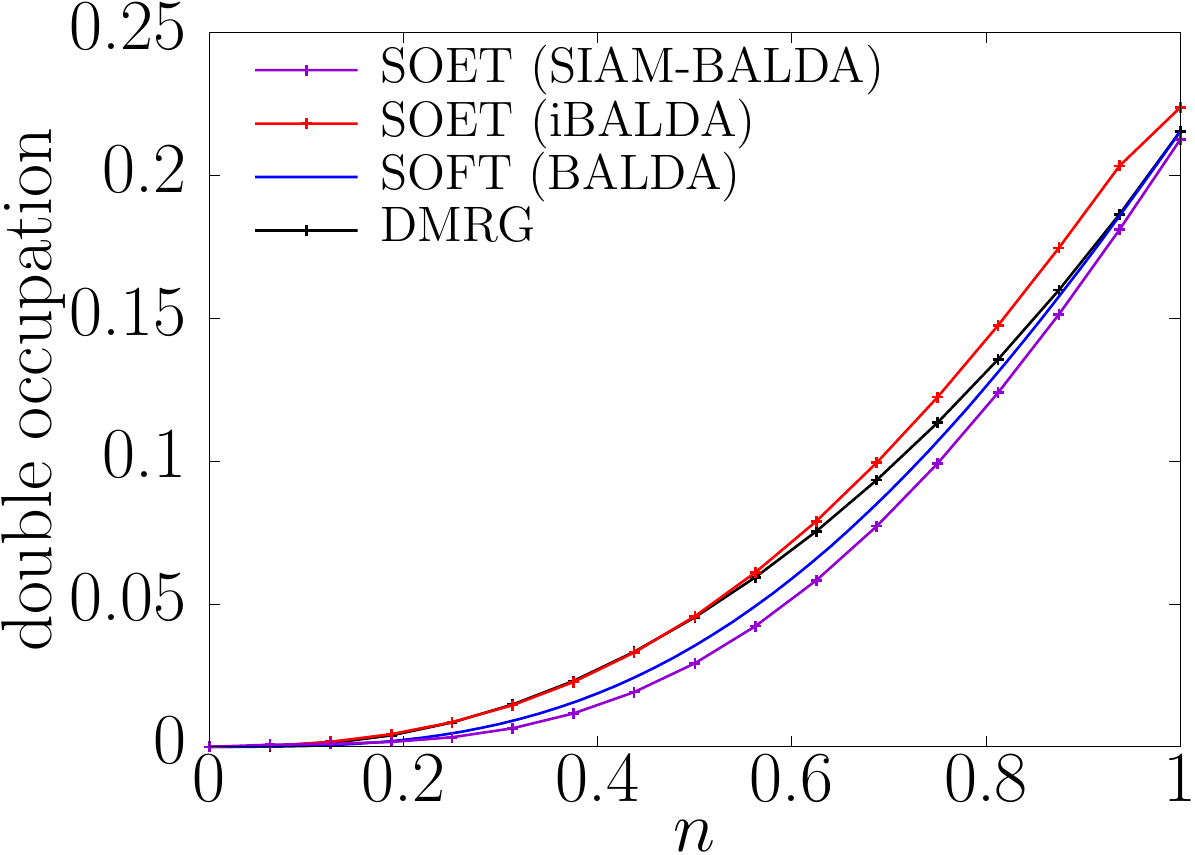}
\caption{Double occupations plotted as a function of the exact density
$n = N/L$ for $U=t=1$ and $L=32$. For iBALDA and SIAM-BALDA, results
obtained with self-consistently converged densities (not shown) are
almost on top of
the ones obtained with the exact densities.}
\label{fig:dblocc_U1}
\end{center}
\end{figure}

Returning to SIAM-BALDA, it gives, for $U = 1$, relatively accurate per-site
energies (see Fig.~\ref{fig:persite-energy}). Interestingly, even though
density-driven errors are still present, they are substantially  
reduced when comparison is made with iBALDA, especially in the low
density regime (see the dashed lines in the top panel of
Fig.~\ref{fig:persite-error}). This is simply due to the fact that,
within SIAM-BALDA, the embedding potential equals the BALDA correlation
potential on {\it all} sites (bath and impurity) and it is complemented
by Yamada's correlation potential (with a minus sign) on the impurity
site only [see Eqs.~(\ref{eq:anderson_yamada}) and
(\ref{eq:Gamma_fun})]. Since the latter potential is quadratic in $U$,
there will be no spurious Hartree contribution to the potential, in
contrast to BALDA (as 
seen from the top panel of Fig.~\ref{fig:Ecimp_wrtn_gammafctn}), and
therefore no self-consistency errors, at least at low density. We also
see from Fig.~\ref{fig:Ecimp_wrtn_gammafctn} that, when the density
increases, the SIAM-BALDA impurity correlation potential, i.e., the
derivative of the SIAM-BALDA impurity correlation energy with respect to $n$, is underestimated (in absolute value) which could explain
why, in this regime, SIAM-BALDA still induces density-driven errors.
We should finally stress that the latter errors might be
enhanced by the fact that we use a complementary per-site correlation functional for
the bath that depends only on the occupation of the impurity (see
Eq.~(\ref{eq:iDFA})). Double occupations are plotted with respect to the
exact density for $U=1$ in Fig.~\ref{fig:dblocc_U1}. As expected,
SIAM-BALDA improves on iBALDA results close to the half-filled case. 
However, at low density, the SIAM-BALDA per-site correlation for the
bath inherits the unphysical linear behavior in
$U$ of BALDA (which is removed in iBALDA by construction), thus leading
to underestimated double occupations. In
summary, the simple version of SIAM-BALDA that we propose is relatively
accurate close to half-filling. By construction, it is in principle only applicable
to relatively
weak correlation regimes. Its generalization to
stronger correlation regimes as well as the possibility to include
occupations of the bath sites in the design of impurity correlation
functionals will be investigated in the future.

\section{Conclusions and perspectives}\label{sec:conclusion}

SOET is an in-principle-exact DFT-based embedding theory where the
(not necessarily uniform) fully-interacting Hubbard system is mapped onto an
impurity-interacting one. 
In this work, SOET has been applied to the uniform one-dimensional
Hubbard model for the purpose of deriving local density approximations.
Exact properties of the embedding functionals have been derived first. In
particular, we have shown that, in order to calculate per-site energies and
double occupations, the contribution of the bath to the
per-site correlation energy is, in addition to the latter, the key
quantity to model in SOET. Various density-functional approximations,
which are based on Bethe ansatz
and perturbative solutions to the Hubbard and Anderson models, have been constructed and
tested. Each functional is well adapted to a particular regime of correlation
and density. For example, one of them (SIAM-BALDA), while performing well around
half-filling and in not too strong correlation regimes, inherits the limitations of BALDA away from half-filling.   
We hope that this work will pave the way to the design of better SOET
density functionals that are applicable to all regimes.

Another key aspect of SOET is the self-consistent calculation of the
embedded impurity system's wavefunction. DMRG has been used for that purpose in this
work but any other wavefunction-based method could in principle be used
in this context.     
In contrast to the DMET, SOET can easily incorporate 
correlation effects in the bath thanks to a dedicated density
functional. 
Note that, for practical
purposes, the SOET
could be reformulated as an 
open impurity site problem, thus allowing for clearer connections
between the two approaches. Work is currently in progress in this
direction.

Let us also mention that the impurity problem in SOET could
alternatively be
solved with a SIAM solver. The use of Green function techniques in SOET is
interesting, not only for practical purposes, but also for the
development of better density-functional embedding potentials. For that
purpose, one would have to derive a Sham--Schl\"{u}ter
equation~\cite{DFT_energygap} for the embedded impurity system. Moreover, a
Green-function-based formulation of SOET might provide a convenient
framework for comparing SOET with DMFT, in the spirit of
Ref.~\cite{ayral2017dynamical}.
Extensions
to higher-dimensional 
~\cite{ijas2010lattice,karlsson2011time,DFT_ModelHamiltonians}
and {\it ab-initio} Hamiltonians [using localized orbitals or
(delocalized) natural molecular orbitals] are also under investigation. 

\begin{acknowledgments}
The authors thank Eli Kraisler, Matthieu Sauban\`{e}re, Bernard Amadon and Lucia Reining for
fruitful 
discussions 
as well as 
Vincent Caudrelier for his
enlightening lectures on the BA.
This work was funded by the Ecole Doctorale des Sciences Chimiques 222 
(Strasbourg), the ANR (MCFUNEX project, Grant No. ANR-14-CE06- 0014-01), the "Japon-Unistra" network as well as the Building of Consortia for the Development of Human Resources in Science and Technology, MEXT, Japan for travel funding.
\end{acknowledgments}

\appendix

\section{\label{sec:dblocc} Exact expression for the double occupation}

In this section, the proof for the SOET-based double occupation expression in 
Eq.~(\ref{eq:doubleocc_soet}) is given.
Let us start with Eq.~(\ref{eq:dblocc_KS}) which is valid for a 
uniform $N$-electron and $L$-site model with density $\underline{n} \equiv \lbrace 
n_i = n \rbrace_i$ where $n = N/L$.
According to Eq.~(\ref{eq:local_ecbath_exp}), the double occupation now reads:
\begin{eqnarray}\label{eq:dblocc_Ehxcimp_ecbath}
d & = & \dfrac{n^2}{4} +  \dfrac{\partial E_{\rm c}^{\rm imp}
(\underline{n})}{\partial U}  + \dfrac{\partial \overline{e}_{\rm 
c}^{\rm bath}(\underline{n})}{\partial U} \nonumber \\
 &= & \dfrac{\partial E_{\rm Hxc}^{\rm imp}(\underline{n})}
 {\partial U}  + \dfrac{\partial \overline{e}_{\rm c}^{\rm bath}
 (\underline{n})}{\partial U}.
\end{eqnarray}
We will come back to this expression later on. Let us now consider
the impurity-interacting LL functional in 
Eq.~(\ref{eq:HK_imp}) which is defined for {\it any} (i.e., not
necessarily uniform) density $\mathbf{n}$.
The minimizing wavefunction in Eq.~(\ref{eq:HK_imp}), whose density
equals $\mathbf{n}$, is denoted 
$\Psi^{\rm imp}(\mathbf{n})$ so that
\begin{eqnarray}\label{eq:Fimp_HK}
F^{\rm imp}(\mathbf{n}) =
\langle \Psi^{\rm imp}(\mathbf{n}) \vert \hat{T} + U 
\hat{n}_{0\uparrow}\hat{n}_{0\downarrow} \vert \Psi^{\rm 
imp}(\mathbf{n}) \rangle.
\end{eqnarray}
An equivalent and useful expression is obtained from the following Legendre-Fenchel 
transform~\cite{fromager2015exact,senjean2017local}:
\begin{eqnarray}\label{eq:LF_Fimp}
F^{\rm imp}(\mathbf{n}) =\sup_{\mathbf{v}} \left \lbrace
\mathcal{E}^{\rm imp}(U,t,\mathbf{v}) - (\mathbf{v} | \mathbf{n}) 
\right \rbrace,
\end{eqnarray}
where $\mathcal{E}^{\rm imp}(U,t,\mathbf{v})$ is the ground-state energy
of $\hat{H}^{\rm imp}(U,t,\mathbf{v})=\hat{T} + \hat{U}_0 + \sum_i v_i \hat{n}_i$.
Differentiating Eq.~(\ref{eq:LF_Fimp}) 
with respect to $U$ gives 
\begin{eqnarray}\label{eq:deriv_LF_sationary_pot}
\dfrac{\partial F^{\rm imp}(\mathbf{n})}{\partial U} = \left. 
\dfrac{\partial \mathcal{E}^{\rm imp}(U,t,\mathbf{v})}{\partial 
U}\right|_{\mathbf{v} = \mathbf{v}^{\rm emb}(U,t,\mathbf{n})},
\end{eqnarray}
where $\mathbf{v}^{\rm emb}(U,t,\mathbf{n})$ is the 
maximizing (and therefore stationary)
potential in Eq.~(\ref{eq:LF_Fimp}), thus leading to the following
expression,
according to the Hellmann-Feynman theorem, 
\begin{eqnarray}\label{eq:dFimp_dU=dimp}
\dfrac{\partial F^{\rm imp}(\mathbf{n})}{\partial U} & = & 
\langle \Psi^{\rm imp}(\mathbf{n}) \vert 
\hat{n}_{0\uparrow}\hat{n}_{0\downarrow} \vert \Psi^{\rm 
imp}(\mathbf{n}) \rangle \nonumber \\ 
& = & d^{\rm imp}(\mathbf{n}),
\end{eqnarray}
where we used the fact that $\Psi^{\rm imp}(\mathbf{n})$, whose double
occupation for the impurity site is denoted $d^{\rm imp}(\mathbf{n})$, is the
ground-state wavefunction of $\hat{H}^{\rm imp}(U,t,\mathbf{v}^{\rm
emb}(U,t,\mathbf{n}))$~\cite{fromager2015exact,senjean2017local}.
Finally, since  
the non-interacting kinetic energy $T_{\rm s}(\mathbf{n})$ does not
depend on $U$, we obtain from Eqs.~(\ref{eq:HK_KS_imp})
and~(\ref{eq:dFimp_dU=dimp}) the general expression 
\begin{eqnarray}\label{eq:dFimp_dU=dEhxcimp_dU}
d^{\rm imp}(\mathbf{n})= \dfrac{\partial E^{\rm imp}_{\rm Hxc}(\mathbf{n})}{\partial U}.
\end{eqnarray}
Returning to the particular case of a uniform $\mathbf{n} =
\underline{n}$ density [see Eq.~(\ref{eq:dblocc_Ehxcimp_ecbath})], it
comes from Eq.~(\ref{eq:dFimp_dU=dEhxcimp_dU}) the following exact
expression for the true physical double occupation,
\begin{eqnarray}\label{eq:doubleocc_soet_appendices}
d& = & d^{\rm imp}(\underline{n}) + \dfrac{\partial 
\overline{e}_{\rm c}^{\rm bath}(\underline{n})}{\partial U},
\end{eqnarray}
which is equivalent to Eq.~(\ref{eq:doubleocc_soet}), since, in the
exact theory, the self-consistent solution $\Psi^{\rm
imp}$ to Eq.~(\ref{eq:self-consistent-SOET}) 
equals (in the uniform case)
 $\Psi^{\rm
imp}(\underline{n})$ and $\underline{n}=\mathbf{n}^{\Psi^{\rm
imp}(\underline{n})}$.

\section{\label{sec:persite} Exact expression for the per-site energy}

In this section, the proof for the SOET per-site energy expression in
Eq.~(\ref{eq:per-site-energy}) is given.
Let us start with Eq.~(\ref{eq:per-site-energy_KS}) which is valid for a 
uniform density profile $\underline{n}$.
According to Eq.~(\ref{eq:local_ecbath_exp}), the per-site energy now reads:
\begin{eqnarray}
e &= & t_{\rm s}(n) + \dfrac{U}{4} n^2  + E_{\rm c}^{\rm imp}(\underline{n}) + \overline{e}_{\rm c}^{\rm bath}(\underline{n}) \nonumber \\
& = & t_{\rm s}(n) + E_{\rm Hxc}^{\rm imp}(\underline{n}) + \overline{e}_{\rm c}^{\rm bath}(\underline{n}),
\end{eqnarray}
or, equivalently, according to Eq.~(\ref{eq:HK_KS_imp}), 
\begin{eqnarray}\label{persite_1}
e & = & t_{\rm s}(n) + F^{\rm imp}(\underline{n}) - T_{\rm s}(\underline{n}) + \overline{e}^{\rm bath}_{\rm c}(\underline{n}).
\end{eqnarray}
We will come back to this relation later on. Let us now focus on the
impurity-interacting LL functional that, for {\it any} (i.e., not
necessarily uniform) density $\mathbf{n}$, we decompose into kinetic
and interaction energy contributions,
\begin{eqnarray}\label{eq:Fimp=Timp+Uimp}
F^{\rm imp}(\mathbf{n}) & = & \langle \Psi^{\rm imp}(\mathbf{n}) \vert
\hat{T} + \hat{U}_0 \vert \Psi^{\rm imp}(\mathbf{n}) \rangle \nonumber \\
& = &  T^{\rm imp}(\mathbf{n}) + U^{\rm imp}(\mathbf{n}),
\end{eqnarray}
where
\begin{eqnarray}
T^{\rm imp}(\mathbf{n}) =  \langle \Psi^{\rm imp}(\mathbf{n}) \vert \hat{T} \vert \Psi^{\rm imp}(\mathbf{n}) \rangle  ,
\end{eqnarray}
and 
\begin{eqnarray}
\label{eq:Uimp_exp}
U^{\rm imp}(\mathbf{n}) & = & \langle \Psi^{\rm imp}(\mathbf{n}) \vert \hat{U}_0 \vert \Psi^{\rm imp}(\mathbf{n}) \rangle \nonumber \\
& = & U  \langle \Psi^{\rm imp}(\mathbf{n}) \vert \hat{n}_{0\uparrow}\hat{n}_{0\downarrow} \vert \Psi^{\rm imp}(\mathbf{n}) \rangle \nonumber \\
 & = & U d^{\rm imp}(\mathbf{n}).
\end{eqnarray}
Following the same strategy as in Eqs.~(\ref{eq:deriv_LF_sationary_pot})
and (\ref{eq:dFimp_dU=dimp}), we obtain
\begin{eqnarray}
\dfrac{\partial F^{\rm imp}(\mathbf{n})}{\partial t} &=& 
\left.\dfrac{\partial 
\mathcal{E}^{\rm imp}(U,t,\mathbf{v})}{\partial t}\right|_{\mathbf{v} = \mathbf{v}^{\rm emb}(U,t,\mathbf{n})}
\nonumber\\
&=&\dfrac{T^{\rm imp}
(\mathbf{n})}{t},
\label{eq:dFimp_dt}
\end{eqnarray}
which gives, when $U=0$, 
\begin{eqnarray}
\dfrac{\partial T_{\rm s}(\mathbf{n})}{\partial t} = 
\dfrac{T_{\rm s}(\mathbf{n})}{t}.
\label{eq:dTs_dt}
\end{eqnarray}
Combining Eqs.~(\ref{eq:HK_KS_imp}), (\ref{eq:Hx_plus_c_for_imp}), (\ref{eq:dFimp_dt}) and 
(\ref{eq:dTs_dt}) finally leads to 
\begin{eqnarray}\label{eq:Timp}
T^{\rm imp}(\mathbf{n}) &=&
t\dfrac{\partial F^{\rm imp}(\mathbf{n})}{\partial t}
\nonumber\\
&=& T_{\rm s}(\mathbf{n}) + t \dfrac{\partial E_{\rm c}^{\rm imp}(\mathbf{n})}{\partial t}.
\end{eqnarray}
Returning to the uniform case $\mathbf{n} = \underline{n}$, it comes
from Eqs.~(\ref{persite_1}), (\ref{eq:Fimp=Timp+Uimp}),
(\ref{eq:Uimp_exp}), and (\ref{eq:Timp}) the following exact expression
for the per-site energy,
\begin{eqnarray}
e & = & t_{\rm s}(n) + Ud^{\rm imp}(\underline{n})+ t \dfrac{\partial E_{\rm c}^{\rm imp}(\underline{n})}{\partial t} + \overline{e}_{\rm c}^{\rm bath}(\underline{n}),
\end{eqnarray}
which, with the decomposition in Eq.~(\ref{eq:local_ecbath_exp}) and the
fact that $\Psi^{\rm imp}(\underline{n})=\Psi^{\rm imp}$, leads
to Eq.~(\ref{eq:per-site-energy}).

\section{\label{sec:particle-hole} Invariance of $\overline{e}_{\rm
c}^{\rm bath}(\mathbf{n})$ under hole-particle symmetry and consequences
for the embedding potential}

Let us consider any density $\mathbf{n}\equiv\lbrace n_i\rbrace_i$ summing up to a number
$N=\sum_in_i$ of
electrons. Under hole-particle symmetry, this density becomes
$(\underline{2} - \mathbf{n})\equiv\lbrace 2-n_i\rbrace_i$ and the
number of electrons equals $2L-N$ where $L$ is the number of sites. We will prove that these two
densities give the {\it same} correlation energy for the impurity. Let
us start with the Legendre--Fenchel transform in Eq.~(\ref{eq:LF_Fimp})
[the $t$- and $U$-dependence of the impurity system's energy is dropped for convenience]
which, under hole-particle symmetry, becomes
\begin{eqnarray}\label{eq:Fimp-hole}
F^{\rm imp}(\underline{2} - \mathbf{n}) =  \underset{\mathbf{v}}{\rm sup} 
\Big\{ &&\mathcal{E}^{{\rm imp},2L-N}(\mathbf{v})  - 2\sum_i v_i
\nonumber\\
&&+ 
(\mathbf{v}|\mathbf{n})\Big\}.
\end{eqnarray}
From the substitution
\begin{eqnarray}\label{eq:subst_pot_LF}
\mathbf{v}\rightarrow-\mathbf{v},
\end{eqnarray}
we obtain the following equivalent expression,
\begin{eqnarray}\label{eq:Fimp-hole_minusv}
F^{\rm imp}(\underline{2} - \mathbf{n}) =  \underset{\mathbf{v}}{\rm sup} 
\Big\{ &&\mathcal{E}^{{\rm imp},2L-N}(-\mathbf{v})  + 2\sum_i v_i
- 
(\mathbf{v}|\mathbf{n})\Big\},\nonumber \\
\end{eqnarray}
where $\mathcal{E}^{{\rm 
imp},2L-N}(-\mathbf{v})$ is the $(2L-N)$-electron ground-state energy of the
impurity-interacting Hamiltonian
\begin{eqnarray}\label{eq:particle_H_minusv}
\hat{H}^{\rm imp}(-\mathbf{v}) & = & -t \sum_{i\sigma} \left( \hat{c}_{i
\sigma}^\dagger \hat{c}_{i+1\sigma} + \mathrm{H.c.}\right) - \sum_{i\sigma}v_i 
\hat{c}_{i\sigma}^\dagger \hat{c}_{i\sigma} \nonumber \\
& & + U  \hat{c}_{0\uparrow}^\dagger \hat{c}_{0\uparrow} 
\hat{c}_{0\downarrow}^\dagger \hat{c}_{0\downarrow}.
\end{eqnarray}
If we now apply the
hole-particle transformation to the creation and annihilation operators,
\begin{eqnarray}
\hat{c}_{i\sigma}^\dagger&\rightarrow&
\hat{b}_{i\sigma}^\dagger=(-1)^i\hat{c}_{i\sigma}
,
\nonumber\\ 
\hat{c}_{i\sigma}&\rightarrow&\hat{b}_{i\sigma}=(-1)^i\hat{c}_{i\sigma}^\dagger
,
\end{eqnarray}
the Hamiltonian in Eq.~(\ref{eq:particle_H_minusv}) becomes
\begin{eqnarray}
\hat{H}^{\rm imp}(-\mathbf{v})
&=&\hat{H}_h^{\rm imp}(\mathbf{v}) - 2\sum_{i}v_i 
\nonumber\\
&&+ U \left(1 - 
 \sum_{\sigma} \hat{b}_{0\sigma}^\dagger \hat{b}_{0\sigma} \right) ,
\end{eqnarray}
where the hole analog of the impurity-interacting Hamiltonian with
arbitrary potential $\mathbf{v}$ reads 
\begin{eqnarray}\label{eq:holeHam_exp}
 \hat{H}_h^{\rm imp}(\mathbf{v}) & = & -t \sum_{i\sigma} \left(\hat{b}_{i
 \sigma}^\dagger \hat{b}_{i+1\sigma} + \mathrm{H.c.}\right) + \sum_{i\sigma}v_i 
 \hat{b}_{i\sigma}^\dagger\hat{b}_{i\sigma} \nonumber \\
& & + U \hat{b}_{0\uparrow}^\dagger\hat{b}_{0\uparrow} 
\hat{b}_{0\downarrow}^\dagger \hat{b}_{0\downarrow}.
\end{eqnarray}
By shifting the potential on the impurity site as follows, 
\begin{eqnarray}\label{eq:shift_pot_LF}
\mathbf{v}\rightarrow\tilde{\mathbf{v}}\equiv\lbrace v_i-U\delta_{i0}\rbrace_i,
\end{eqnarray}
we finally obtain 
\begin{eqnarray}\label{eq:holeHam_and_shift}
\hat{H}^{\rm imp}(-\mathbf{v})
&=&\hat{H}_h^{\rm imp}(\tilde{\mathbf{v}}) - 2\sum_{i}v_i 
+U.
\end{eqnarray}
As readily seen from Eq.~(\ref{eq:holeHam_and_shift}), 
$\hat{H}^{\rm imp}(-\mathbf{v})$ 
and 
$\hat{H}_h^{\rm
imp}(\tilde{\mathbf{v}})$ 
share the
same $(2L-N)$-electron ground state. Moreover, it is clear from
Eqs.~(\ref{eq:particle_H_minusv}) and (\ref{eq:holeHam_exp})   
that the $(2L-N)$-electron [i.e., $N$-hole] ground-state energy of $\hat{H}_h^{\rm
imp}(\tilde{\mathbf{v}})$ is nothing but the $N$-electron
ground-state energy $\mathcal{E}^{{\rm imp},N}(\tilde{\mathbf{v}})$ of
$\hat{H}^{\rm imp}(\tilde{\mathbf{v}})$. From these observations and Eq.~(\ref{eq:holeHam_and_shift}) we
conclude that  
\begin{eqnarray}\label{eq:impE-N-2L-Nrelation}
\mathcal{E}^{{\rm imp},2L-N}(-\mathbf{v}) & = & \mathcal{E}^{{\rm imp},N}
(\tilde{\mathbf{v}}) - 2 \sum_i v_i +U.
\end{eqnarray}
Introducing Eq.~(\ref{eq:impE-N-2L-Nrelation}) into
Eq.~(\ref{eq:Fimp-hole_minusv}) leads to
\begin{eqnarray}\label{eq:Fimp_2-n_n}
F^{\rm imp}(\underline{2} - \mathbf{n}) & = & 
\underset{\mathbf{v}}{\rm sup} 
\left \lbrace \mathcal{E}^{{\rm imp},N}(\tilde{\mathbf{v}})  
-(\mathbf{v}|\mathbf{n})\right  \rbrace +U\nonumber  \\
&=&
\underset{\tilde{\mathbf{v}}}{\rm sup} 
\left \lbrace \mathcal{E}^{{\rm imp},N}(\tilde{\mathbf{v}})  
-(\tilde{\mathbf{v}}|\mathbf{n})\right  \rbrace +U(1-n_0)\nonumber  \\
&=& F^{{\rm imp}}(\mathbf{n})+U(1-n_0).
\end{eqnarray}
In the particular case $U=0$, we recover the hole-particle symmetry
relation for the non-interacting kinetic energy,
\begin{eqnarray}\label{eq:hp_sym_Ts}
T_{\rm 
s}(\underline{2} - \mathbf{n})
=T_{\rm s}(\mathbf{n}).
\end{eqnarray}
We conclude from Eqs.~(\ref{eq:Fimp_2-n_n}),
(\ref{eq:hp_sym_Ts}),~(\ref{eq:HK_KS_imp}) and (\ref{eq:Hx_plus_c_for_imp}) that
the impurity correlation density-functional energy is invariant under
hole-particle symmetry,  
\begin{eqnarray}
E_{\rm c}^{\rm imp}(\underline{2}-\mathbf{n}) = E_{\rm c}^{\rm imp}
(\mathbf{n}).
\end{eqnarray}
Since the per-site correlation energy in the uniform system is also
invariant~\cite{DFT_ModelHamiltonians}, 
\begin{eqnarray}
e_{\rm c}(2-n)=e_{\rm c}(n),
\end{eqnarray} 
it comes from Eq.~(\ref{eq:local_ecbath_exp}) that $\overline{e}_{\rm c}^{\rm
bath}(\mathbf{n})$ is invariant under hole-particle symmetry.  
As a result, we have
\begin{eqnarray}
-\left.\dfrac{\partial e_{\rm c}(\nu)}{\partial
\nu}\right|_{\nu=2-n}=
\frac{\partial e_{\rm c}(n)}{\partial n},
\end{eqnarray} 
and 
\begin{eqnarray}
-\left.\dfrac{\partial \overline{e}_{\rm c}^{\rm bath}({\bm \nu})}{\partial
\nu_i}\right|_{{\bm \nu}=\underline{2}-\mathbf{n}}=
\dfrac{\partial
\overline{e}_{\rm c}^{\rm bath}(\bf{n})}{\partial n_i},
\end{eqnarray}
which, for a half-filled {\it finite-size} system gives
\begin{eqnarray}\label{eq:zero_pot_half_filling}
\left.\frac{\partial e_{\rm c}(n)}{\partial n}\right|_{n=1}=0=
\left.\dfrac{\partial
\overline{e}_{\rm c}^{\rm bath}(\bf{n})}{\partial n_i}\right|_{{\bf
n}=\underline{1}}.
\end{eqnarray}
Note that, for a finite $L$ value, uniform densities will have discrete values
which explains why we do not distinguish
${\bf{n}}\rightarrow\underline{1}^+$ and
${\bf{n}}\rightarrow\underline{1}^-$ limits and just consider
${\bf{n}}=\underline{1}$. However, in the thermodynamic limit
($L\rightarrow+\infty$), this distinction should be made otherwise 
the physical band gap cannot be reproduced~\cite{lima2002density}. Note also that, for a {\it
finite-size} system with uniform (discrete) density profiles ${\bf n}=\underline{n}$, the maximizing
embedding potential in
Eq.~(\ref{eq:Fimp-hole}) fulfills, according to
Eqs.~(\ref{eq:subst_pot_LF}),~(\ref{eq:Fimp-hole_minusv}), (\ref{eq:shift_pot_LF}) and
(\ref{eq:Fimp_2-n_n}), the following hole-particle symmetry relation
\begin{eqnarray}
{v}_i^{\rm
emb}(\underline{2}-\underline{n})
=-{v}^{\rm emb}_i(\underline{n})
-U\delta_{i0},
\end{eqnarray}    
thus leading to, at half-filling,
\begin{eqnarray}
{v}^{\rm emb}_i(\underline{1})=- \dfrac{U}{2}\delta_{i0}.
\end{eqnarray} 
The latter result is also recovered from
Eq.~(\ref{eq:zero_pot_half_filling}) and Eq.~(\ref{eq:emb_pot_uniform}).

\section{Derivative discontinuity in KS SOFT and SOET at $n=1$ in the atomic limit}\label{sec:derivative discontinuity}

Let us consider the fully-interacting $L$-site Hubbard Hamiltonian in
the atomic limit ($t=0$), 
\begin{eqnarray}
\hat{H}(n)=
\hat{U}+v(n)\sum_i\hat{n}_i
,
\end{eqnarray}
which reproduces the uniform density profile with density $n$. 
Starting from the half-filling situation ($L$
electrons or, equivalently, $n=1$), we can add an electron in order to
investigate the behavior of $v(n)$ when $n\rightarrow1^+$. 
 In order to have a total number of electrons
varying continuously from $L$ to $L+1$ or, equivalently, $1<n<(L+1)/L$,
the $L$- and $(L+1)$-electron ground states of $\hat{H}(n)$ must be
degenerate, thus leading to the following condition,    
\begin{eqnarray}
Lv(n)=(L+1)v(n)+U,
\end{eqnarray}
or, equivalently, $v(n)=-U$. Therefore we conclude that, in the
thermodynamic limit ($L\rightarrow+\infty$),
\begin{eqnarray}
\left.
v(n)
\right|_{n=1^+}=-U.
\end{eqnarray}
On the other hand, if we consider the removal of an electron, the
density can vary continuously in the range $(L-1)/L<n<1$ if the $(L-1)$-
and $L$-electron ground states of $\hat{H}(n)$ are degenerate, thus
leading to the condition $(L-1)v(n)=Lv(n)$ and, consequently,    
\begin{eqnarray}
\left.
v(n)
\right|_{n=1^-}=0.
\end{eqnarray}
Turning to the KS Hamiltonian with ground-state uniform density $n$ (and
$t=0$),  
\begin{eqnarray}
\hat{H}^{\rm KS}(n)=
v^{\rm KS}(n)\sum_i\hat{n}_i,
\end{eqnarray}
we can show similarly that, in contrast to the interacting case,
the KS potential has no discontinuity at $n=1$,
\begin{eqnarray}
\left.
v^{\rm KS}(n)
\right|_{n=1^+}
=
\left.
v^{\rm KS}(n)
\right|_{n=1^-}
=0.
\end{eqnarray}
Consequently, we recover the well-known discontinuous behavior of the
correlation potential at half-filling:    
\begin{eqnarray}\label{eq:dis_vc_1+}
\left.\dfrac{\partial e_{\rm c}(n)}{\partial n}\right|_{n=1^+}
&=&
\left.\left(v^{\rm KS}(n)-v(n)-\dfrac{U}{2}n\right)\right|_{n=1^+}
\nonumber\\
&=&+U/2,
\end{eqnarray}
and
\begin{eqnarray}\label{eq:dis_vc_1-}
\left.\dfrac{\partial e_{\rm c}(n)}{\partial n}\right|_{n=1^-}&=&
\left.\left(v^{\rm KS}(n)-v(n)-\dfrac{U}{2}n\right)\right|_{n=1^-}
\nonumber\\
&=&-U/2.
\end{eqnarray}
Let us now consider the impurity-interacting Hamiltonian 
of SOET in the atomic limit,  
\begin{eqnarray}
\hat{H}^{\rm imp}(n)=
\hat{U}_0+\sum_iv^{\rm emb}_i({n})\hat{n}_i
.
\end{eqnarray}
A uniform
density $n=(L+1)/L$ is obtained from the $(L+1)$-electron ground state of
$\hat{H}^{\rm imp}(n)$ if, for {\it any} bath site label $i$ ($i\neq 0$), 
\begin{eqnarray}\label{eq:deg_cond_L+1}
&&U+2
v^{\rm emb}_0(n)
+
\sum_{k\neq 0}v^{\rm emb}_k(n)
=
2v^{\rm emb}_i(n)
+
\sum_{k\neq i}v^{\rm emb}_k(n)
. \nonumber \\
\end{eqnarray}     
The latter degeneracy
condition simply ensures that the added electron can occupy either the
impurity site or a bath site. In order to let $n$ vary continuously in the
range $1<n<(L+1)/L$ we {\it also} need the $L$- and $(L+1)$-electron ground
states to be degenerate:
\begin{eqnarray}
2v^{\rm emb}_i(n)
+
\sum_{k\neq i}v^{\rm emb}_k(n)
=\sum_{k}v^{\rm emb}_k(n),
\end{eqnarray}
thus leading to $v^{\rm emb}_i(n)=0$ in the bath and, according to
Eq.~(\ref{eq:deg_cond_L+1}), $v^{\rm emb}_0(n)=-U$. If, on the other
hand, the density varies in the range $(L-1)/L<n<1$, the degeneracy
condition between $(L-1)$- and $L$-electron ground states reads
\begin{eqnarray}
\sum_{k\neq j}v^{\rm emb}_k(n)
=\sum_{k}v^{\rm emb}_k(n).
\end{eqnarray}
Note that the latter condition, which leads to $v^{\rm emb}_j(n)=0$, holds for any site (impurity or bath)
label $j$. In summary, we obtain in the thermodynamic limit,
\begin{eqnarray}
\left.
v^{\rm emb}_0(n)
\right|_{n=1^-}&=&0,
\nonumber\\
\left.v^{\rm emb}_0(n)
\right|_{n=1^+}&=&-U,
\nonumber\\
\left.
v^{\rm emb}_i(n)
\right|_{n=1^-}
&=&
\left.
v^{\rm emb}_i(n)
\right|_{n=1^+}
=0,\hspace{0.2cm} i\neq 0.
\end{eqnarray}
We then conclude from Eqs.~(\ref{eq:self-consistent-SOET}),
(\ref{eq:decomp_Hx_c_bath}), and (\ref{eq:final_exp_Ec_bath}) that, in the atomic limit, the complementary
per-site bath correlation potential exhibits no discontinuous behavior
at half-filling, neither on the impurity
site, 
\begin{eqnarray}\label{eq:no_disc_vcbath}
\left.\dfrac{\partial \overline{e}_{\rm
c}^{\rm bath}(\mathbf{n})}{\partial
n_0}\right|_{\mathbf{n}={\underline{1}^+}}
&=&
\left.\Big(v^{\rm
emb}_0(n)-v(n)\Big)\right|_{n=1^+} 
\nonumber\\
&=&0
\nonumber\\
&=&
\left.\Big(v^{\rm
emb}_0(n)-v(n)\Big)\right|_{n=1^-} 
\nonumber\\
&=&
\left.\dfrac{\partial \overline{e}_{\rm
c}^{\rm bath}(\mathbf{n})}{\partial
n_0}\right|_{\mathbf{n}={\underline{1}^-}},
\end{eqnarray}
nor on the bath sites ($i\neq
0$), since
\begin{eqnarray}
\left.\dfrac{\partial \overline{e}_{\rm
c}^{\rm bath}(\mathbf{n})}{\partial
n_i}\right|_{\mathbf{n}={\underline{1}^+}}
&=&
\left.\left(v^{\rm
emb}_i(n)-v(n)-\dfrac{U}{2}n-\dfrac{\partial e_{\rm c}(n)}{\partial
n}\right)\right|_{n=1^+} 
\nonumber\\
&=&\left.\left(v^{\rm
emb}_i(n)-v^{\rm KS}(n)\right)\right|_{n=1^+} 
\nonumber\\
&=&0
\nonumber\\
&=&
\left.\left(v^{\rm
emb}_i(n)-v^{\rm KS}(n)\right)\right|_{n=1^-} 
\nonumber\\
&=&
\left.\dfrac{\partial \overline{e}_{\rm
c}^{\rm bath}(\mathbf{n})}{\partial
n_i}\right|_{\mathbf{n}={\underline{1}^-}}.
\end{eqnarray}
Note that, as a consequence of Eqs.~(\ref{eq:dis_vc_1+}), (\ref{eq:dis_vc_1-}), and (\ref{eq:no_disc_vcbath}),
\begin{eqnarray}
\left.\dfrac{\partial E_{\rm c}^{\rm imp}({\bf n})}{\partial
n_0}\right|_{\mathbf{n}={\underline{1}^+}}
&=&
\left.\dfrac{\partial e_{\rm c}(n_0)}{\partial n_0}\right|_{n_0=1^+}
-
\left.\dfrac{\partial \overline{e}_{\rm
c}^{\rm bath}(\mathbf{n})}{\partial
n_0}\right|_{\mathbf{n}={\underline{1}^+}}
\nonumber\\
&=&+U/2,
\end{eqnarray}
and
\begin{eqnarray}
\left.\dfrac{\partial E_{\rm c}^{\rm imp}({\bf n})}{\partial
n_0}\right|_{\mathbf{n}={\underline{1}^-}}
&=&
\left.\dfrac{\partial e_{\rm c}(n_0)}{\partial n_0}\right|_{n_0=1^-}
-
\left.\dfrac{\partial \overline{e}_{\rm
c}^{\rm bath}(\mathbf{n})}{\partial
n_0}\right|_{\mathbf{n}={\underline{1}^-}}
\nonumber\\
&=&-U/2. \\
\nonumber
\end{eqnarray}
Therefore, the impurity correlation potential exhibits a discontinuity at half-filling.


\begin{thebibliography}{77}
\expandafter\ifx\csname natexlab\endcsname\relax\def\natexlab#1{#1}\fi
\expandafter\ifx\csname bibnamefont\endcsname\relax
  \def\bibnamefont#1{#1}\fi
\expandafter\ifx\csname bibfnamefont\endcsname\relax
  \def\bibfnamefont#1{#1}\fi
\expandafter\ifx\csname citenamefont\endcsname\relax
  \def\citenamefont#1{#1}\fi
\expandafter\ifx\csname url\endcsname\relax
  \def\url#1{\texttt{#1}}\fi
\expandafter\ifx\csname urlprefix\endcsname\relax\def\urlprefix{URL }\fi
\providecommand{\bibinfo}[2]{#2}
\providecommand{\eprint}[2][]{\url{#2}}

\bibitem[{\citenamefont{Anisimov et~al.}(1991)\citenamefont{Anisimov, Zaanen,
  and Andersen}}]{anisimov1991band}
\bibinfo{author}{\bibfnamefont{V.~I.} \bibnamefont{Anisimov}},
  \bibinfo{author}{\bibfnamefont{J.}~\bibnamefont{Zaanen}}, \bibnamefont{and}
  \bibinfo{author}{\bibfnamefont{O.~K.} \bibnamefont{Andersen}},
  \bibinfo{journal}{Phys. Rev. B} \textbf{\bibinfo{volume}{44}},
  \bibinfo{pages}{943} (\bibinfo{year}{1991}).

\bibitem[{\citenamefont{Liechtenstein et~al.}(1995)\citenamefont{Liechtenstein,
  Anisimov, and Zaanen}}]{liechtenstein1995density}
\bibinfo{author}{\bibfnamefont{A.}~\bibnamefont{Liechtenstein}},
  \bibinfo{author}{\bibfnamefont{V.}~\bibnamefont{Anisimov}}, \bibnamefont{and}
  \bibinfo{author}{\bibfnamefont{J.}~\bibnamefont{Zaanen}},
  \bibinfo{journal}{Phys. Rev. B} \textbf{\bibinfo{volume}{52}},
  \bibinfo{pages}{R5467} (\bibinfo{year}{1995}).

\bibitem[{\citenamefont{Pulay}(1983)}]{pulay1983localizability}
\bibinfo{author}{\bibfnamefont{P.}~\bibnamefont{Pulay}},
  \bibinfo{journal}{Chem. Phys. Lett.} \textbf{\bibinfo{volume}{100}},
  \bibinfo{pages}{151} (\bibinfo{year}{1983}).

\bibitem[{\citenamefont{Saebo and Pulay}(1993)}]{saebo1993local}
\bibinfo{author}{\bibfnamefont{S.}~\bibnamefont{Saebo}} \bibnamefont{and}
  \bibinfo{author}{\bibfnamefont{P.}~\bibnamefont{Pulay}},
  \bibinfo{journal}{Annu. Rev. Phys. Chem.} \textbf{\bibinfo{volume}{44}},
  \bibinfo{pages}{213} (\bibinfo{year}{1993}).

\bibitem[{\citenamefont{Hampel and Werner}(1996)}]{hampel1996local}
\bibinfo{author}{\bibfnamefont{C.}~\bibnamefont{Hampel}} \bibnamefont{and}
  \bibinfo{author}{\bibfnamefont{H.-J.} \bibnamefont{Werner}},
  \bibinfo{journal}{J. Chem. Phys.} \textbf{\bibinfo{volume}{104}},
  \bibinfo{pages}{6286} (\bibinfo{year}{1996}).

\bibitem[{\citenamefont{Sun and Chan}(2016)}]{sun2016quantum}
\bibinfo{author}{\bibfnamefont{Q.}~\bibnamefont{Sun}} \bibnamefont{and}
  \bibinfo{author}{\bibfnamefont{G.~K.-L.} \bibnamefont{Chan}},
  \bibinfo{journal}{Acc. Chem. Res.} \textbf{\bibinfo{volume}{49}},
  \bibinfo{pages}{2705} (\bibinfo{year}{2016}).

\bibitem[{\citenamefont{Ayral et~al.}(2017)\citenamefont{Ayral, Lee, and
  Kotliar}}]{ayral2017dynamical}
\bibinfo{author}{\bibfnamefont{T.}~\bibnamefont{Ayral}},
  \bibinfo{author}{\bibfnamefont{T.-H.} \bibnamefont{Lee}}, \bibnamefont{and}
  \bibinfo{author}{\bibfnamefont{G.}~\bibnamefont{Kotliar}},
  \bibinfo{journal}{Phys. Rev. B} \textbf{\bibinfo{volume}{96}},
  \bibinfo{pages}{235139} (\bibinfo{year}{2017}).

\bibitem[{\citenamefont{Georges and Kotliar}(1992)}]{georges1992hubbard}
\bibinfo{author}{\bibfnamefont{A.}~\bibnamefont{Georges}} \bibnamefont{and}
  \bibinfo{author}{\bibfnamefont{G.}~\bibnamefont{Kotliar}},
  \bibinfo{journal}{Phys. Rev. B} \textbf{\bibinfo{volume}{45}},
  \bibinfo{pages}{6479} (\bibinfo{year}{1992}).

\bibitem[{\citenamefont{Georges et~al.}(1996)\citenamefont{Georges, Kotliar,
  Krauth, and Rozenberg}}]{georges1996limitdimension}
\bibinfo{author}{\bibfnamefont{A.}~\bibnamefont{Georges}},
  \bibinfo{author}{\bibfnamefont{G.}~\bibnamefont{Kotliar}},
  \bibinfo{author}{\bibfnamefont{W.}~\bibnamefont{Krauth}}, \bibnamefont{and}
  \bibinfo{author}{\bibfnamefont{M.~J.} \bibnamefont{Rozenberg}},
  \bibinfo{journal}{Rev. Mod. Phys.} \textbf{\bibinfo{volume}{68}}
  (\bibinfo{year}{1996}).

\bibitem[{\citenamefont{Kotliar and Vollhardt}(2004)}]{kotliar2004strongly}
\bibinfo{author}{\bibfnamefont{G.}~\bibnamefont{Kotliar}} \bibnamefont{and}
  \bibinfo{author}{\bibfnamefont{D.}~\bibnamefont{Vollhardt}},
  \bibinfo{journal}{Phys. Today} \textbf{\bibinfo{volume}{57}},
  \bibinfo{pages}{53} (\bibinfo{year}{2004}).

\bibitem[{\citenamefont{Held}(2007)}]{held2007electronic}
\bibinfo{author}{\bibfnamefont{K.}~\bibnamefont{Held}}, \bibinfo{journal}{Adv.
  Phys.} \textbf{\bibinfo{volume}{56}}, \bibinfo{pages}{829}
  (\bibinfo{year}{2007}).

\bibitem[{\citenamefont{Zgid and Chan}(2011)}]{zgid2011DMFTquantum}
\bibinfo{author}{\bibfnamefont{D.}~\bibnamefont{Zgid}} \bibnamefont{and}
  \bibinfo{author}{\bibfnamefont{G.~K.-L.} \bibnamefont{Chan}},
  \bibinfo{journal}{J. Chem. Phys.} \textbf{\bibinfo{volume}{134}}
  (\bibinfo{year}{2011}).

\bibitem[{\citenamefont{Kotliar et~al.}(2006)\citenamefont{Kotliar, Savrasov,
  Haule, Oudovenko, Parcollet, and Marianetti}}]{kotliar2006reviewDMFT}
\bibinfo{author}{\bibfnamefont{G.}~\bibnamefont{Kotliar}},
  \bibinfo{author}{\bibfnamefont{S.~Y.} \bibnamefont{Savrasov}},
  \bibinfo{author}{\bibfnamefont{K.}~\bibnamefont{Haule}},
  \bibinfo{author}{\bibfnamefont{V.~S.} \bibnamefont{Oudovenko}},
  \bibinfo{author}{\bibfnamefont{O.}~\bibnamefont{Parcollet}},
  \bibnamefont{and} \bibinfo{author}{\bibfnamefont{C.~A.}
  \bibnamefont{Marianetti}}, \bibinfo{journal}{Rev. Mod. Phys.}
  \textbf{\bibinfo{volume}{78}}, \bibinfo{pages}{865} (\bibinfo{year}{2006}).

\bibitem[{\citenamefont{Sun and Kotliar}(2002)}]{sun2002extended}
\bibinfo{author}{\bibfnamefont{P.}~\bibnamefont{Sun}} \bibnamefont{and}
  \bibinfo{author}{\bibfnamefont{G.}~\bibnamefont{Kotliar}},
  \bibinfo{journal}{Phys. Rev. B} \textbf{\bibinfo{volume}{66}},
  \bibinfo{pages}{085120} (\bibinfo{year}{2002}).

\bibitem[{\citenamefont{Biermann et~al.}(2003)\citenamefont{Biermann,
  Aryasetiawan, and Georges}}]{biermann2003first}
\bibinfo{author}{\bibfnamefont{S.}~\bibnamefont{Biermann}},
  \bibinfo{author}{\bibfnamefont{F.}~\bibnamefont{Aryasetiawan}},
  \bibnamefont{and} \bibinfo{author}{\bibfnamefont{A.}~\bibnamefont{Georges}},
  \bibinfo{journal}{Phys. Rev. Lett.} \textbf{\bibinfo{volume}{90}},
  \bibinfo{pages}{086402} (\bibinfo{year}{2003}).

\bibitem[{\citenamefont{Karlsson}(2005)}]{karlsson2005self}
\bibinfo{author}{\bibfnamefont{K.}~\bibnamefont{Karlsson}},
  \bibinfo{journal}{J. Phys. Condens. Matter} \textbf{\bibinfo{volume}{17}},
  \bibinfo{pages}{7573} (\bibinfo{year}{2005}).

\bibitem[{\citenamefont{Boehnke et~al.}(2016)\citenamefont{Boehnke, Nilsson,
  Aryasetiawan, and Werner}}]{boehnke2016strong}
\bibinfo{author}{\bibfnamefont{L.}~\bibnamefont{Boehnke}},
  \bibinfo{author}{\bibfnamefont{F.}~\bibnamefont{Nilsson}},
  \bibinfo{author}{\bibfnamefont{F.}~\bibnamefont{Aryasetiawan}},
  \bibnamefont{and} \bibinfo{author}{\bibfnamefont{P.}~\bibnamefont{Werner}},
  \bibinfo{journal}{Phys. Rev. B} \textbf{\bibinfo{volume}{94}},
  \bibinfo{pages}{201106} (\bibinfo{year}{2016}).

\bibitem[{\citenamefont{Werner and Casula}(2016)}]{werner2016dynamical}
\bibinfo{author}{\bibfnamefont{P.}~\bibnamefont{Werner}} \bibnamefont{and}
  \bibinfo{author}{\bibfnamefont{M.}~\bibnamefont{Casula}},
  \bibinfo{journal}{J. Phys. Condens. Matter} \textbf{\bibinfo{volume}{28}},
  \bibinfo{pages}{383001} (\bibinfo{year}{2016}).

\bibitem[{\citenamefont{Nilsson et~al.}(2017)\citenamefont{Nilsson, Boehnke,
  Werner, and Aryasetiawan}}]{nilsson2017multitier}
\bibinfo{author}{\bibfnamefont{F.}~\bibnamefont{Nilsson}},
  \bibinfo{author}{\bibfnamefont{L.}~\bibnamefont{Boehnke}},
  \bibinfo{author}{\bibfnamefont{P.}~\bibnamefont{Werner}}, \bibnamefont{and}
  \bibinfo{author}{\bibfnamefont{F.}~\bibnamefont{Aryasetiawan}},
  \bibinfo{journal}{Phys. Rev. Materials} \textbf{\bibinfo{volume}{1}},
  \bibinfo{pages}{043803} (\bibinfo{year}{2017}).

\bibitem[{\citenamefont{Kananenka et~al.}(2015)\citenamefont{Kananenka, Gull,
  and Zgid}}]{kananenka2015systematically}
\bibinfo{author}{\bibfnamefont{A.~A.} \bibnamefont{Kananenka}},
  \bibinfo{author}{\bibfnamefont{E.}~\bibnamefont{Gull}}, \bibnamefont{and}
  \bibinfo{author}{\bibfnamefont{D.}~\bibnamefont{Zgid}},
  \bibinfo{journal}{Phys. Rev. B} \textbf{\bibinfo{volume}{91}},
  \bibinfo{pages}{121111} (\bibinfo{year}{2015}).

\bibitem[{\citenamefont{Lan et~al.}(2015)\citenamefont{Lan, Kananenka, and
  Zgid}}]{lan2015communication}
\bibinfo{author}{\bibfnamefont{T.~N.} \bibnamefont{Lan}},
  \bibinfo{author}{\bibfnamefont{A.~A.} \bibnamefont{Kananenka}},
  \bibnamefont{and} \bibinfo{author}{\bibfnamefont{D.}~\bibnamefont{Zgid}},
  \bibinfo{journal}{J. Chem. Phys.} \textbf{\bibinfo{volume}{143}},
  \bibinfo{pages}{241102} (\bibinfo{year}{2015}).

\bibitem[{\citenamefont{Lan et~al.}(2017)\citenamefont{Lan, Shee, Li, Gull, and
  Zgid}}]{lan2017testing}
\bibinfo{author}{\bibfnamefont{T.~N.} \bibnamefont{Lan}},
  \bibinfo{author}{\bibfnamefont{A.}~\bibnamefont{Shee}},
  \bibinfo{author}{\bibfnamefont{J.}~\bibnamefont{Li}},
  \bibinfo{author}{\bibfnamefont{E.}~\bibnamefont{Gull}}, \bibnamefont{and}
  \bibinfo{author}{\bibfnamefont{D.}~\bibnamefont{Zgid}},
  \bibinfo{journal}{Phys. Rev. B} \textbf{\bibinfo{volume}{96}},
  \bibinfo{pages}{155106} (\bibinfo{year}{2017}).

\bibitem[{\citenamefont{Knizia and Chan}(2012)}]{knizia2012density}
\bibinfo{author}{\bibfnamefont{G.}~\bibnamefont{Knizia}} \bibnamefont{and}
  \bibinfo{author}{\bibfnamefont{G.~K.-L.} \bibnamefont{Chan}},
  \bibinfo{journal}{Phys. Rev. Lett.} \textbf{\bibinfo{volume}{109}},
  \bibinfo{pages}{186404} (\bibinfo{year}{2012}).

\bibitem[{\citenamefont{Knizia and Chan}(2013)}]{knizia2013density}
\bibinfo{author}{\bibfnamefont{G.}~\bibnamefont{Knizia}} \bibnamefont{and}
  \bibinfo{author}{\bibfnamefont{G.~K.-L.} \bibnamefont{Chan}},
  \bibinfo{journal}{J. Chem. Theory Comput.} \textbf{\bibinfo{volume}{9}},
  \bibinfo{pages}{1428} (\bibinfo{year}{2013}).

\bibitem[{\citenamefont{Bulik et~al.}(2014)\citenamefont{Bulik, Scuseria, and
  Dukelsky}}]{bulik2014density}
\bibinfo{author}{\bibfnamefont{I.~W.} \bibnamefont{Bulik}},
  \bibinfo{author}{\bibfnamefont{G.~E.} \bibnamefont{Scuseria}},
  \bibnamefont{and} \bibinfo{author}{\bibfnamefont{J.}~\bibnamefont{Dukelsky}},
  \bibinfo{journal}{Phys. Rev. B} \textbf{\bibinfo{volume}{89}},
  \bibinfo{pages}{035140} (\bibinfo{year}{2014}).

\bibitem[{\citenamefont{Zheng and Chan}(2016)}]{zheng2016ground}
\bibinfo{author}{\bibfnamefont{B.-X.} \bibnamefont{Zheng}} \bibnamefont{and}
  \bibinfo{author}{\bibfnamefont{G.~K.-L.} \bibnamefont{Chan}},
  \bibinfo{journal}{Phys. Rev. B} \textbf{\bibinfo{volume}{93}},
  \bibinfo{pages}{035126} (\bibinfo{year}{2016}).

\bibitem[{\citenamefont{Wouters
  et~al.}(2016{\natexlab{a}})\citenamefont{Wouters, Jim{\'e}nez-Hoyos, Sun, and
  Chan}}]{wouters2016practical}
\bibinfo{author}{\bibfnamefont{S.}~\bibnamefont{Wouters}},
  \bibinfo{author}{\bibfnamefont{C.~A.} \bibnamefont{Jim{\'e}nez-Hoyos}},
  \bibinfo{author}{\bibfnamefont{Q.}~\bibnamefont{Sun}}, \bibnamefont{and}
  \bibinfo{author}{\bibfnamefont{G.~K.-L.} \bibnamefont{Chan}},
  \bibinfo{journal}{J. Chem. Theory Comput.}
  (\bibinfo{year}{2016}{\natexlab{a}}).

\bibitem[{\citenamefont{Wouters
  et~al.}(2016{\natexlab{b}})\citenamefont{Wouters, Jim{\'e}nez-Hoyos, and
  Chan}}]{wouters2016five}
\bibinfo{author}{\bibfnamefont{S.}~\bibnamefont{Wouters}},
  \bibinfo{author}{\bibfnamefont{C.~A.} \bibnamefont{Jim{\'e}nez-Hoyos}},
  \bibnamefont{and} \bibinfo{author}{\bibfnamefont{G.~K.-L.}
  \bibnamefont{Chan}}, \bibinfo{journal}{arXiv preprint arXiv:1605.05547}
  (\bibinfo{year}{2016}{\natexlab{b}}).

\bibitem[{\citenamefont{Rubin}(2016)}]{rubin2016hybrid}
\bibinfo{author}{\bibfnamefont{N.~C.} \bibnamefont{Rubin}},
  \bibinfo{journal}{arXiv preprint arXiv:1610.06910}  (\bibinfo{year}{2016}).

\bibitem[{\citenamefont{Tsuchimochi et~al.}(2015)\citenamefont{Tsuchimochi,
  Welborn, and Van~Voorhis}}]{tsuchimochi2015density}
\bibinfo{author}{\bibfnamefont{T.}~\bibnamefont{Tsuchimochi}},
  \bibinfo{author}{\bibfnamefont{M.}~\bibnamefont{Welborn}}, \bibnamefont{and}
  \bibinfo{author}{\bibfnamefont{T.}~\bibnamefont{Van~Voorhis}},
  \bibinfo{journal}{J. Chem. Phys.} \textbf{\bibinfo{volume}{143}},
  \bibinfo{pages}{024107} (\bibinfo{year}{2015}).

\bibitem[{\citenamefont{Welborn et~al.}(2016)\citenamefont{Welborn,
  Tsuchimochi, and Van~Voorhis}}]{welborn2016bootstrap}
\bibinfo{author}{\bibfnamefont{M.}~\bibnamefont{Welborn}},
  \bibinfo{author}{\bibfnamefont{T.}~\bibnamefont{Tsuchimochi}},
  \bibnamefont{and}
  \bibinfo{author}{\bibfnamefont{T.}~\bibnamefont{Van~Voorhis}},
  \bibinfo{journal}{J. Chem. Phys.} \textbf{\bibinfo{volume}{145}},
  \bibinfo{pages}{074102} (\bibinfo{year}{2016}).

\bibitem[{\citenamefont{Chayes et~al.}(1985)\citenamefont{Chayes, Chayes, and
  Ruskai}}]{chayes1985density}
\bibinfo{author}{\bibfnamefont{J.}~\bibnamefont{Chayes}},
  \bibinfo{author}{\bibfnamefont{L.}~\bibnamefont{Chayes}}, \bibnamefont{and}
  \bibinfo{author}{\bibfnamefont{M.~B.} \bibnamefont{Ruskai}},
  \bibinfo{journal}{J. Stat. Phys.} \textbf{\bibinfo{volume}{38}},
  \bibinfo{pages}{497} (\bibinfo{year}{1985}).

\bibitem[{\citenamefont{Gunnarsson and
  Sch{\"o}nhammer}(1986)}]{gunnarsson1986density}
\bibinfo{author}{\bibfnamefont{O.}~\bibnamefont{Gunnarsson}} \bibnamefont{and}
  \bibinfo{author}{\bibfnamefont{K.}~\bibnamefont{Sch{\"o}nhammer}},
  \bibinfo{journal}{Phys. Rev. Lett.} \textbf{\bibinfo{volume}{56}},
  \bibinfo{pages}{1968} (\bibinfo{year}{1986}).

\bibitem[{\citenamefont{Sch\"onhammer et~al.}(1995)\citenamefont{Sch\"onhammer,
  Gunnarsson, and Noack}}]{DFT_lattice}
\bibinfo{author}{\bibfnamefont{K.}~\bibnamefont{Sch\"onhammer}},
  \bibinfo{author}{\bibfnamefont{O.}~\bibnamefont{Gunnarsson}},
  \bibnamefont{and} \bibinfo{author}{\bibfnamefont{R.}~\bibnamefont{Noack}},
  \bibinfo{journal}{Phys. Rev. B} \textbf{\bibinfo{volume}{52}},
  \bibinfo{pages}{2504} (\bibinfo{year}{1995}).

\bibitem[{\citenamefont{Capelle and {Campo Jr.}}(2013)}]{DFT_ModelHamiltonians}
\bibinfo{author}{\bibfnamefont{K.}~\bibnamefont{Capelle}} \bibnamefont{and}
  \bibinfo{author}{\bibfnamefont{V.~L.} \bibnamefont{{Campo Jr.}}},
  \bibinfo{journal}{Phys. Rep.} \textbf{\bibinfo{volume}{528}},
  \bibinfo{pages}{91} (\bibinfo{year}{2013}).

\bibitem[{\citenamefont{Lima et~al.}(2002)\citenamefont{Lima, Oliveira, and
  Capelle}}]{lima2002density}
\bibinfo{author}{\bibfnamefont{N.}~\bibnamefont{Lima}},
  \bibinfo{author}{\bibfnamefont{L.}~\bibnamefont{Oliveira}}, \bibnamefont{and}
  \bibinfo{author}{\bibfnamefont{K.}~\bibnamefont{Capelle}},
  \bibinfo{journal}{Europhys. Lett.} \textbf{\bibinfo{volume}{60}},
  \bibinfo{pages}{601} (\bibinfo{year}{2002}).

\bibitem[{\citenamefont{Lima et~al.}(2003)\citenamefont{Lima, Silva, Oliveira,
  and Capelle}}]{lima2003density}
\bibinfo{author}{\bibfnamefont{N.~A.} \bibnamefont{Lima}},
  \bibinfo{author}{\bibfnamefont{M.~F.} \bibnamefont{Silva}},
  \bibinfo{author}{\bibfnamefont{L.~N.} \bibnamefont{Oliveira}},
  \bibnamefont{and} \bibinfo{author}{\bibfnamefont{K.}~\bibnamefont{Capelle}},
  \bibinfo{journal}{Phys. Rev. Lett.} \textbf{\bibinfo{volume}{90}},
  \bibinfo{pages}{146402} (\bibinfo{year}{2003}).

\bibitem[{\citenamefont{Capelle et~al.}(2003)\citenamefont{Capelle, Lima,
  Silva, and Oliveira}}]{capelle2003density}
\bibinfo{author}{\bibfnamefont{K.}~\bibnamefont{Capelle}},
  \bibinfo{author}{\bibfnamefont{N.}~\bibnamefont{Lima}},
  \bibinfo{author}{\bibfnamefont{M.}~\bibnamefont{Silva}}, \bibnamefont{and}
  \bibinfo{author}{\bibfnamefont{L.}~\bibnamefont{Oliveira}}, in
  \emph{\bibinfo{booktitle}{The Fundamentals of Electron Density, Density
  Matrix and Density Functional Theory in Atoms, Molecules and the Solid
  State}} (\bibinfo{publisher}{Springer}, \bibinfo{year}{2003}), p.
  \bibinfo{pages}{145}.

\bibitem[{\citenamefont{Franca et~al.}(2012)\citenamefont{Franca, Vieira, and
  Capelle}}]{franca2012simple}
\bibinfo{author}{\bibfnamefont{V.~V.} \bibnamefont{Franca}},
  \bibinfo{author}{\bibfnamefont{D.}~\bibnamefont{Vieira}}, \bibnamefont{and}
  \bibinfo{author}{\bibfnamefont{K.}~\bibnamefont{Capelle}},
  \bibinfo{journal}{New J. Phys.} \textbf{\bibinfo{volume}{14}},
  \bibinfo{pages}{073021} (\bibinfo{year}{2012}).

\bibitem[{\citenamefont{Xianlong et~al.}(2006)\citenamefont{Xianlong, Polini,
  Tosi, Campo~Jr, Capelle, and Rigol}}]{xianlong2006bethe}
\bibinfo{author}{\bibfnamefont{G.}~\bibnamefont{Xianlong}},
  \bibinfo{author}{\bibfnamefont{M.}~\bibnamefont{Polini}},
  \bibinfo{author}{\bibfnamefont{M.}~\bibnamefont{Tosi}},
  \bibinfo{author}{\bibfnamefont{V.~L.} \bibnamefont{Campo~Jr}},
  \bibinfo{author}{\bibfnamefont{K.}~\bibnamefont{Capelle}}, \bibnamefont{and}
  \bibinfo{author}{\bibfnamefont{M.}~\bibnamefont{Rigol}},
  \bibinfo{journal}{Phys. Rev. B} \textbf{\bibinfo{volume}{73}},
  \bibinfo{pages}{165120} (\bibinfo{year}{2006}).

\bibitem[{\citenamefont{Xianlong et~al.}(2007)\citenamefont{Xianlong, Rizzi,
  Polini, Fazio, Tosi, Campo~Jr, and Capelle}}]{xianlong2007luther}
\bibinfo{author}{\bibfnamefont{G.}~\bibnamefont{Xianlong}},
  \bibinfo{author}{\bibfnamefont{M.}~\bibnamefont{Rizzi}},
  \bibinfo{author}{\bibfnamefont{M.}~\bibnamefont{Polini}},
  \bibinfo{author}{\bibfnamefont{R.}~\bibnamefont{Fazio}},
  \bibinfo{author}{\bibfnamefont{M.~P.} \bibnamefont{Tosi}},
  \bibinfo{author}{\bibfnamefont{V.~L.} \bibnamefont{Campo~Jr}},
  \bibnamefont{and} \bibinfo{author}{\bibfnamefont{K.}~\bibnamefont{Capelle}},
  \bibinfo{journal}{Phys. Rev. Lett.} \textbf{\bibinfo{volume}{98}},
  \bibinfo{pages}{030404} (\bibinfo{year}{2007}).

\bibitem[{\citenamefont{Silva et~al.}(2005)\citenamefont{Silva, Lima, Malvezzi,
  and Capelle}}]{silva2005effects}
\bibinfo{author}{\bibfnamefont{M.}~\bibnamefont{Silva}},
  \bibinfo{author}{\bibfnamefont{N.}~\bibnamefont{Lima}},
  \bibinfo{author}{\bibfnamefont{A.~L.} \bibnamefont{Malvezzi}},
  \bibnamefont{and} \bibinfo{author}{\bibfnamefont{K.}~\bibnamefont{Capelle}},
  \bibinfo{journal}{Phys. Rev. B} \textbf{\bibinfo{volume}{71}},
  \bibinfo{pages}{125130} (\bibinfo{year}{2005}).

\bibitem[{\citenamefont{Akande and Sanvito}(2010)}]{akande2010electric}
\bibinfo{author}{\bibfnamefont{A.}~\bibnamefont{Akande}} \bibnamefont{and}
  \bibinfo{author}{\bibfnamefont{S.}~\bibnamefont{Sanvito}},
  \bibinfo{journal}{Phys. Rev. B} \textbf{\bibinfo{volume}{82}},
  \bibinfo{pages}{245114} (\bibinfo{year}{2010}).

\bibitem[{\citenamefont{Campo~Jr and Capelle}(2005)}]{campo2005phase}
\bibinfo{author}{\bibfnamefont{V.}~\bibnamefont{Campo~Jr}} \bibnamefont{and}
  \bibinfo{author}{\bibfnamefont{K.}~\bibnamefont{Capelle}},
  \bibinfo{journal}{Phys. Rev. A} \textbf{\bibinfo{volume}{72}},
  \bibinfo{pages}{061602} (\bibinfo{year}{2005}).

\bibitem[{\citenamefont{Xianlong}(2008)}]{xianlong2008effects}
\bibinfo{author}{\bibfnamefont{G.}~\bibnamefont{Xianlong}},
  \bibinfo{journal}{Phys. Rev. B} \textbf{\bibinfo{volume}{78}},
  \bibinfo{pages}{085108} (\bibinfo{year}{2008}).

\bibitem[{\citenamefont{Bergfield et~al.}(2012)\citenamefont{Bergfield, Liu,
  Burke, and Stafford}}]{bergfield2012bethe}
\bibinfo{author}{\bibfnamefont{J.~P.} \bibnamefont{Bergfield}},
  \bibinfo{author}{\bibfnamefont{Z.-F.} \bibnamefont{Liu}},
  \bibinfo{author}{\bibfnamefont{K.}~\bibnamefont{Burke}}, \bibnamefont{and}
  \bibinfo{author}{\bibfnamefont{C.~A.} \bibnamefont{Stafford}},
  \bibinfo{journal}{Phys. Rev. Lett.} \textbf{\bibinfo{volume}{108}},
  \bibinfo{pages}{066801} (\bibinfo{year}{2012}).

\bibitem[{\citenamefont{Liu et~al.}(2012)\citenamefont{Liu, Bergfield, Burke,
  and Stafford}}]{liu2012accuracy}
\bibinfo{author}{\bibfnamefont{Z.-F.} \bibnamefont{Liu}},
  \bibinfo{author}{\bibfnamefont{J.~P.} \bibnamefont{Bergfield}},
  \bibinfo{author}{\bibfnamefont{K.}~\bibnamefont{Burke}}, \bibnamefont{and}
  \bibinfo{author}{\bibfnamefont{C.~A.} \bibnamefont{Stafford}},
  \bibinfo{journal}{Phys. Rev. B} \textbf{\bibinfo{volume}{85}},
  \bibinfo{pages}{155117} (\bibinfo{year}{2012}).

\bibitem[{\citenamefont{Liu and Burke}(2015)}]{liu2015coulomb}
\bibinfo{author}{\bibfnamefont{Z.-F.} \bibnamefont{Liu}} \bibnamefont{and}
  \bibinfo{author}{\bibfnamefont{K.}~\bibnamefont{Burke}},
  \bibinfo{journal}{Phys. Rev. B} \textbf{\bibinfo{volume}{91}},
  \bibinfo{pages}{245158} (\bibinfo{year}{2015}).

\bibitem[{\citenamefont{Verdozzi}(2008)}]{verdozzi2008time}
\bibinfo{author}{\bibfnamefont{C.}~\bibnamefont{Verdozzi}},
  \bibinfo{journal}{Phys. Rev. Lett.} \textbf{\bibinfo{volume}{101}},
  \bibinfo{pages}{166401} (\bibinfo{year}{2008}).

\bibitem[{\citenamefont{Kurth et~al.}(2010)\citenamefont{Kurth, Stefanucci,
  Khosravi, Verdozzi, and Gross}}]{kurth2010dynamical}
\bibinfo{author}{\bibfnamefont{S.}~\bibnamefont{Kurth}},
  \bibinfo{author}{\bibfnamefont{G.}~\bibnamefont{Stefanucci}},
  \bibinfo{author}{\bibfnamefont{E.}~\bibnamefont{Khosravi}},
  \bibinfo{author}{\bibfnamefont{C.}~\bibnamefont{Verdozzi}}, \bibnamefont{and}
  \bibinfo{author}{\bibfnamefont{E.}~\bibnamefont{Gross}},
  \bibinfo{journal}{Phys. Rev. Lett.} \textbf{\bibinfo{volume}{104}},
  \bibinfo{pages}{236801} (\bibinfo{year}{2010}).

\bibitem[{\citenamefont{Karlsson et~al.}(2011)\citenamefont{Karlsson,
  Privitera, and Verdozzi}}]{karlsson2011time}
\bibinfo{author}{\bibfnamefont{D.}~\bibnamefont{Karlsson}},
  \bibinfo{author}{\bibfnamefont{A.}~\bibnamefont{Privitera}},
  \bibnamefont{and} \bibinfo{author}{\bibfnamefont{C.}~\bibnamefont{Verdozzi}},
  \bibinfo{journal}{Phys. Rev. Lett.} \textbf{\bibinfo{volume}{106}},
  \bibinfo{pages}{116401} (\bibinfo{year}{2011}).

\bibitem[{\citenamefont{Stefanucci and Kurth}(2011)}]{stefanucci2011towards}
\bibinfo{author}{\bibfnamefont{G.}~\bibnamefont{Stefanucci}} \bibnamefont{and}
  \bibinfo{author}{\bibfnamefont{S.}~\bibnamefont{Kurth}},
  \bibinfo{journal}{Phys. Rev. Lett.} \textbf{\bibinfo{volume}{107}},
  \bibinfo{pages}{216401} (\bibinfo{year}{2011}).

\bibitem[{\citenamefont{Xianlong et~al.}(2012)\citenamefont{Xianlong, Chen,
  Tokatly, and Kurth}}]{xianlong2012lattice}
\bibinfo{author}{\bibfnamefont{G.}~\bibnamefont{Xianlong}},
  \bibinfo{author}{\bibfnamefont{A.-H.} \bibnamefont{Chen}},
  \bibinfo{author}{\bibfnamefont{I.}~\bibnamefont{Tokatly}}, \bibnamefont{and}
  \bibinfo{author}{\bibfnamefont{S.}~\bibnamefont{Kurth}},
  \bibinfo{journal}{Phys. Rev. B} \textbf{\bibinfo{volume}{86}},
  \bibinfo{pages}{235139} (\bibinfo{year}{2012}).

\bibitem[{\citenamefont{Schindlmayr and Godby}(1995)}]{schindlmayr1995density}
\bibinfo{author}{\bibfnamefont{A.}~\bibnamefont{Schindlmayr}} \bibnamefont{and}
  \bibinfo{author}{\bibfnamefont{R.}~\bibnamefont{Godby}},
  \bibinfo{journal}{Phys. Rev. B} \textbf{\bibinfo{volume}{51}},
  \bibinfo{pages}{10427} (\bibinfo{year}{1995}).

\bibitem[{\citenamefont{L{\'o}pez-Sandoval and
  Pastor}(2002)}]{lopez2002density}
\bibinfo{author}{\bibfnamefont{R.}~\bibnamefont{L{\'o}pez-Sandoval}}
  \bibnamefont{and} \bibinfo{author}{\bibfnamefont{G.}~\bibnamefont{Pastor}},
  \bibinfo{journal}{Phys. Rev. B} \textbf{\bibinfo{volume}{66}},
  \bibinfo{pages}{155118} (\bibinfo{year}{2002}).

\bibitem[{\citenamefont{L{\'o}pez-Sandoval and
  Pastor}(2004)}]{lopez2004interaction}
\bibinfo{author}{\bibfnamefont{R.}~\bibnamefont{L{\'o}pez-Sandoval}}
  \bibnamefont{and} \bibinfo{author}{\bibfnamefont{G.}~\bibnamefont{Pastor}},
  \bibinfo{journal}{Phys. Rev. B} \textbf{\bibinfo{volume}{69}},
  \bibinfo{pages}{085101} (\bibinfo{year}{2004}).

\bibitem[{\citenamefont{Sauban{\`e}re and Pastor}(2009)}]{saubanere2009scaling}
\bibinfo{author}{\bibfnamefont{M.}~\bibnamefont{Sauban{\`e}re}}
  \bibnamefont{and} \bibinfo{author}{\bibfnamefont{G.}~\bibnamefont{Pastor}},
  \bibinfo{journal}{Phys. Rev. B} \textbf{\bibinfo{volume}{79}},
  \bibinfo{pages}{235101} (\bibinfo{year}{2009}).

\bibitem[{\citenamefont{Sauban{\`e}re and Pastor}(2011)}]{saubanere2011density}
\bibinfo{author}{\bibfnamefont{M.}~\bibnamefont{Sauban{\`e}re}}
  \bibnamefont{and} \bibinfo{author}{\bibfnamefont{G.}~\bibnamefont{Pastor}},
  \bibinfo{journal}{Phys. Rev. B} \textbf{\bibinfo{volume}{84}},
  \bibinfo{pages}{035111} (\bibinfo{year}{2011}).

\bibitem[{\citenamefont{T{\"o}ws and Pastor}(2011)}]{tows2011lattice}
\bibinfo{author}{\bibfnamefont{W.}~\bibnamefont{T{\"o}ws}} \bibnamefont{and}
  \bibinfo{author}{\bibfnamefont{G.}~\bibnamefont{Pastor}},
  \bibinfo{journal}{Phys. Rev. B} \textbf{\bibinfo{volume}{83}},
  \bibinfo{pages}{235101} (\bibinfo{year}{2011}).

\bibitem[{\citenamefont{T{\"o}ws et~al.}(2014)\citenamefont{T{\"o}ws,
  Sauban{\`e}re, and Pastor}}]{tows2014density}
\bibinfo{author}{\bibfnamefont{W.}~\bibnamefont{T{\"o}ws}},
  \bibinfo{author}{\bibfnamefont{M.}~\bibnamefont{Sauban{\`e}re}},
  \bibnamefont{and} \bibinfo{author}{\bibfnamefont{G.}~\bibnamefont{Pastor}},
  \bibinfo{journal}{Theor. Chem. Acc.} \textbf{\bibinfo{volume}{133}},
  \bibinfo{pages}{1} (\bibinfo{year}{2014}).

\bibitem[{\citenamefont{Sauban{\`e}re et~al.}(2016)\citenamefont{Sauban{\`e}re,
  Lepetit, and Pastor}}]{saubanere2016interaction}
\bibinfo{author}{\bibfnamefont{M.}~\bibnamefont{Sauban{\`e}re}},
  \bibinfo{author}{\bibfnamefont{M.~B.} \bibnamefont{Lepetit}},
  \bibnamefont{and} \bibinfo{author}{\bibfnamefont{G.}~\bibnamefont{Pastor}},
  \bibinfo{journal}{Phys. Rev. B} \textbf{\bibinfo{volume}{94}},
  \bibinfo{pages}{045102} (\bibinfo{year}{2016}).

\bibitem[{\citenamefont{Stefanucci and Kurth}(2015)}]{stefanucci2015steady}
\bibinfo{author}{\bibfnamefont{G.}~\bibnamefont{Stefanucci}} \bibnamefont{and}
  \bibinfo{author}{\bibfnamefont{S.}~\bibnamefont{Kurth}},
  \bibinfo{journal}{Nano Lett.} \textbf{\bibinfo{volume}{15}},
  \bibinfo{pages}{8020} (\bibinfo{year}{2015}).

\bibitem[{\citenamefont{Kurth and Stefanucci}(2016)}]{kurth2016nonequilibrium}
\bibinfo{author}{\bibfnamefont{S.}~\bibnamefont{Kurth}} \bibnamefont{and}
  \bibinfo{author}{\bibfnamefont{G.}~\bibnamefont{Stefanucci}},
  \bibinfo{journal}{Phys. Rev. B} \textbf{\bibinfo{volume}{94}},
  \bibinfo{pages}{241103} (\bibinfo{year}{2016}).

\bibitem[{\citenamefont{Fromager}(2015)}]{fromager2015exact}
\bibinfo{author}{\bibfnamefont{E.}~\bibnamefont{Fromager}},
  \bibinfo{journal}{Mol. Phys.} \textbf{\bibinfo{volume}{113}},
  \bibinfo{pages}{419} (\bibinfo{year}{2015}).

\bibitem[{\citenamefont{Senjean et~al.}(2017)\citenamefont{Senjean, Tsuchiizu,
  Robert, and Fromager}}]{senjean2017local}
\bibinfo{author}{\bibfnamefont{B.}~\bibnamefont{Senjean}},
  \bibinfo{author}{\bibfnamefont{M.}~\bibnamefont{Tsuchiizu}},
  \bibinfo{author}{\bibfnamefont{V.}~\bibnamefont{Robert}}, \bibnamefont{and}
  \bibinfo{author}{\bibfnamefont{E.}~\bibnamefont{Fromager}},
  \bibinfo{journal}{Mol. Phys.} \textbf{\bibinfo{volume}{115}},
  \bibinfo{pages}{48} (\bibinfo{year}{2017}).

\bibitem[{\citenamefont{White}(1992)}]{white1992density}
\bibinfo{author}{\bibfnamefont{S.~R.} \bibnamefont{White}},
  \bibinfo{journal}{Phys. Rev. Lett.} \textbf{\bibinfo{volume}{69}},
  \bibinfo{pages}{2863} (\bibinfo{year}{1992}).

\bibitem[{\citenamefont{White}(1993)}]{white1993density}
\bibinfo{author}{\bibfnamefont{S.~R.} \bibnamefont{White}},
  \bibinfo{journal}{Phys. Rev. B} \textbf{\bibinfo{volume}{48}},
  \bibinfo{pages}{10345} (\bibinfo{year}{1993}).

\bibitem[{\citenamefont{Verstraete et~al.}(2008)\citenamefont{Verstraete, Murg,
  and Cirac}}]{verstraete2008matrix}
\bibinfo{author}{\bibfnamefont{F.}~\bibnamefont{Verstraete}},
  \bibinfo{author}{\bibfnamefont{V.}~\bibnamefont{Murg}}, \bibnamefont{and}
  \bibinfo{author}{\bibfnamefont{J.~I.} \bibnamefont{Cirac}},
  \bibinfo{journal}{Adv. Phys.} \textbf{\bibinfo{volume}{57}},
  \bibinfo{pages}{143} (\bibinfo{year}{2008}).

\bibitem[{\citenamefont{Schollw{\"o}ck}(2011)}]{schollwock2011density}
\bibinfo{author}{\bibfnamefont{U.}~\bibnamefont{Schollw{\"o}ck}},
  \bibinfo{journal}{Ann. Phys.} \textbf{\bibinfo{volume}{326}},
  \bibinfo{pages}{96} (\bibinfo{year}{2011}).

\bibitem[{\citenamefont{Nakatani}()}]{naokicode}
\bibinfo{author}{\bibfnamefont{N.}~\bibnamefont{Nakatani}},
  \bibinfo{howpublished}{{[Online]. https://github.com/naokin/mpsxx}}.

\bibitem[{\citenamefont{Lieb and Wu}(1968)}]{NoMott_Hubbardmodel}
\bibinfo{author}{\bibfnamefont{E.~H.} \bibnamefont{Lieb}} \bibnamefont{and}
  \bibinfo{author}{\bibfnamefont{F.~Y.} \bibnamefont{Wu}},
  \bibinfo{journal}{Phys. Rev. Lett.} \textbf{\bibinfo{volume}{20}},
  \bibinfo{pages}{1445} (\bibinfo{year}{1968}).

\bibitem[{\citenamefont{Anderson}(1961)}]{Anderson}
\bibinfo{author}{\bibfnamefont{P.~W.} \bibnamefont{Anderson}},
  \bibinfo{journal}{Phys. Rev.} \textbf{\bibinfo{volume}{124}},
  \bibinfo{pages}{1} (\bibinfo{year}{1961}).

\bibitem[{\citenamefont{Ying et~al.}(2014)\citenamefont{Ying, Brosco, and
  Lorenzana}}]{ying2014solving}
\bibinfo{author}{\bibfnamefont{Z.-J.} \bibnamefont{Ying}},
  \bibinfo{author}{\bibfnamefont{V.}~\bibnamefont{Brosco}}, \bibnamefont{and}
  \bibinfo{author}{\bibfnamefont{J.}~\bibnamefont{Lorenzana}},
  \bibinfo{journal}{Phys. Rev. B} \textbf{\bibinfo{volume}{89}},
  \bibinfo{pages}{205130} (\bibinfo{year}{2014}).

\bibitem[{\citenamefont{Georges}(2016)}]{CRP16_Georges_impurities}
\bibinfo{author}{\bibfnamefont{A.}~\bibnamefont{Georges}}, \bibinfo{journal}{C.
  R. Phys.} \textbf{\bibinfo{volume}{17}}, \bibinfo{pages}{430 }
  (\bibinfo{year}{2016}).

\bibitem[{\citenamefont{Yamada}(1975)}]{yamada1975perturbation}
\bibinfo{author}{\bibfnamefont{K.}~\bibnamefont{Yamada}},
  \bibinfo{journal}{Prog. Theo. Phys.} \textbf{\bibinfo{volume}{53}},
  \bibinfo{pages}{970} (\bibinfo{year}{1975}).

\bibitem[{\citenamefont{Sham and Schl\"{u}ter}(1983)}]{DFT_energygap}
\bibinfo{author}{\bibfnamefont{L.~J.} \bibnamefont{Sham}} \bibnamefont{and}
  \bibinfo{author}{\bibfnamefont{M.}~\bibnamefont{Schl\"{u}ter}},
  \bibinfo{journal}{Phys. Rev. B} \textbf{\bibinfo{volume}{51}},
  \bibinfo{pages}{1888} (\bibinfo{year}{1983}).

\bibitem[{\citenamefont{Ij{\"a}s and Harju}(2010)}]{ijas2010lattice}
\bibinfo{author}{\bibfnamefont{M.}~\bibnamefont{Ij{\"a}s}} \bibnamefont{and}
  \bibinfo{author}{\bibfnamefont{A.}~\bibnamefont{Harju}},
  \bibinfo{journal}{Phys. Rev. B} \textbf{\bibinfo{volume}{82}},
  \bibinfo{pages}{235111} (\bibinfo{year}{2010}).

\end{thebibliography}
\end{document}